\documentclass[a4paper,11pt]{article}
\pdfoutput=1
\usepackage[T1]{fontenc}
\usepackage{xcolor}
\usepackage{jheppub}

\usepackage[normalem]{ulem}

\usepackage{bm}
\usepackage{mathtools}
\usepackage{epstopdf}
\usepackage{setspace}
\usepackage{dsfont}
\usepackage{latexsym}
\usepackage{psfrag}
\usepackage{booktabs}
\usepackage{color}
\usepackage{subfig}
\usepackage{tensor}
\usepackage{lscape}
\usepackage{tikz}
\usepackage[inline]{enumitem}

\usepackage{cleveref}

\usetikzlibrary{shapes}
\usetikzlibrary{patterns}
\usetikzlibrary{matrix,arrows.meta} 				
\usetikzlibrary{positioning}				
\usetikzlibrary{calc,through}				
\usetikzlibrary{decorations.pathreplacing}  
\usetikzlibrary{decorations.pathmorphing}	

\pgfmathsetmacro{\h}{1.5}
\pgfmathsetmacro{\hp}{2}
\pgfmathsetmacro{\w}{1.5}
\pgfmathsetmacro{\blb}{0.25}
\pgfmathsetmacro{\ar}{0.2}

\usetikzlibrary{decorations.markings}
\tikzset{
    fermion/.style={draw=black, postaction={decorate},
        decoration={markings,mark=at position .55 with {\arrow[draw=black]{>}}}},
    fermionbar/.style={draw=black, postaction={decorate},
        decoration={markings,mark=at position .55 with {\arrow[draw=black]{<}}}},
    graviton/.style={decorate, draw=black,
        decoration={coil,aspect=0,amplitude=1.5pt, segment length=5pt}}
}

\tikzstyle{block} = [draw, rectangle, 
    minimum height=3em, minimum width=6earticlem]

\definecolor{darkblue}{rgb}{0.2, 0, 0.8}

\renewcommand{\r}{\rho}

\newcommand{\abs}[1]{\left\lvert #1 \right\rvert}

\definecolor{cadmiumgreen}{rgb}{0.0, 0.42, 0.24}

\def\be{\begin{equation}}
\def\ee{\end{equation}}
\def\bea{\begin{eqnarray}}
\def\eea{\end{eqnarray}}
\def\ba{\begin{array}}
\def\ea{\end{array}}
\def\bd{\begin{displaymath}}
\def\ed{\end{displaymath}}

\def\b{\beta}

\def\r{\rho}                                     

\def\>{\rangle} 
\def\<{\langle} 
\def\to{\rightarrow}

\def\dd{\textrm{d}}
\def\r{|\vec{x}|}

\def\hdelta{\hat{\delta}}

\newcommand{\id}{1\hspace{-3pt}{\rm l}}

\def\kk{\mathsf{k}}

\def\htilde{{\widetilde h}}
\def\Ctilde{{\widetilde C}}
\def\Psitilde{{\widetilde \Psi}}

\crefname{figure}{figure}{figures}
\Crefname{figure}{Figure}{Figures}

\title{Asymptotic Simplicity 
and Scattering in General Relativity from Quantum Field Theory}

\author[a]{Stefano De Angelis,}
\author[b]{Aidan Herderschee,}
\author[c,d,e]{Radu Roiban}
\author[f,d]{and  Fei Teng}

\affiliation[a]{Institut de Physique Th\'eorique, CEA, CNRS, Universit\'e Paris-Saclay, \\
    F-91191 Gif-sur-Yvette cedex, France}
\affiliation[b]{Institute for Advanced Study, 
    Princeton, NJ 08540, USA}
\affiliation[c]{Institute for Theoretical Studies, ETH Zurich, 8092 Zurich, Switzerland}    
\affiliation[d]{Department of Physics, Pennsylvania State University, University Park, PA 16802, USA}
\affiliation[e]{Institute for Computational and Data Sciences,
Pennsylvania State University,
University Park, PA 16802, USA}
\affiliation[f]{Department of Physics and Center for Field Theory and Particle Physics, Fudan University, Shanghai 200438, China}

\emailAdd{stefano.de-angelis@ipht.fr}
\emailAdd{aidanh@ias.edu}
\emailAdd{radu@phys.psu.edu}
\emailAdd{f\_teng@fudan.edu.cn}

\abstract{
    We investigate the fate of asymptotic simplicity in physically relevant settings of compact-object scattering. Using the stress tensor of a two-body system as a source, we compute the spacetime metric in General Relativity at finite observer distance in an asymptotic expansion. To do so, we relate the metric to the final-state graviton one-point function in momentum space, which is computed using perturbative QFT techniques.
    Both the simple pole and the infrared-related logarithmic branch cut in the virtuality of the external graviton contribute nontrivially.
    We focus on determining the fall-off behavior of the Newman-Penrose scalars, confirming previous predictions that Sachs's peeling property is violated at leading order in the post-Minkowski expansion. 
    Our analysis at higher orders in the post-Minkowskian expansion reveals a significantly stronger breakdown of the peeling property than previously recognized, which is the result of nonlinear, long-range interactions between localized sources 
    and the surrounding gravitational field.
}

\begin{document}

\addtocontents{toc}{\protect\setcounter{tocdepth}{2}}

\maketitle


\section{Introduction}
\label{sec:intro}

The spacetime sourced by an isolated physical system is expected to approach Minkowski spacetime asymptotically. The latter can be conformally compactified \cite{Penrose:1965am}, becoming a space 
with smooth future and past null infinities, ${\cal I}^+$ and ${\cal I}^-$ respectively. It is therefore natural to expect that the spacetime of isolated physical systems admits a conformal compactification. That is, it can be expected that there exists a conformal factor such that the physical spacetime, once rescaled, can be isometrically embedded into the unphysical one and extended smoothly to its null boundaries. 
Distilling extensive earlier work, see e.g.~~\cite{Bondi:1960jsa,Bondi:1962px,Bondi:1962rkt,Sachs:1961zz,Sachs:1962wk,Newman:1961qr,Penrose:1962ij,Penrose:1964ge}, 
Penrose introduced the notion of asymptotic simplicity as a formalization of this idea~\cite{Penrose:1965am} and conjectured that it should be a property of all physically-relevant spacetimes. In turn, this implies that all physically-relevant spacetimes have a conformal compactification with $C^{4}$ regularity~\cite{Friedrich:2017cjg}.
See Refs.~\cite{Valiente-Kroon:2002xys,Kroon:2004me,Kehrberger:2021uvf,Kehrberger:2021vhp,Kehrberger:2021azo,Gajic:2022pst,Kehrberger:2024clh,Kehrberger:2024aak,Geiller:2024ryw} for recent developments. 
It has been found that large classes of initial data do preserve asymptotic simplicity, see Refs.~\cite{Blanchet:1985sp,Blanchet:1986dk,friedrich1983cauchy,friedrich1986existence,andersson1992regularity,chrusciel2002existence,Corvino:2007}. 

The asymptotic structure of spacetime at null infinity plays an important, if understated, role in many modern developments in quantum field theory, being essential for identifying asymptotic symmetries and for defining a scattering theory for massless fields.
For example, it has been shown that the infrared behavior of scattering amplitudes in both gauge and gravity theories are constrained by soft theorems of spontaneously-broken asymptotic symmetries~\cite{He:2014laa,Campiglia:2014yka,Cachazo:2014fwa,Strominger:2017zoo}. 
More radically, it has been proposed that the holographic principle \cite{tHooft:1993dmi,Susskind:1994vu} admits a realization in asymptotically flat spacetimes such that suitably-defined infrared-finite S-matrix elements are given by certain correlation functions in a dual theory defined at null infinity~\cite{Pasterski:2016qvg,Pasterski:2017kqt,Laddha:2020kvp,Guevara:2021abz,Raclariu:2021zjz,Pasterski:2021rjz,Pasterski:2021raf,Costello:2022wso}. 
Thus, in spacetimes in which asymptotic simplicity were violated in a sufficiently strong manner, the associated asymptotic symmetries may be altered, 
warranting a reexamination of some of these conclusions.
This should not pose difficulties in calculations where the departure 
from asymptotic simplicity is irrelevant, such as the computation of 
IR finite observables around Minkowski space.

Direct verification of the $C^4$ regularity of the conformally compactified metric is typically computationally prohibitive. A more practical approach is to infer it by analyzing other consequences of the assumption of asymptotic simplicity. One of them is the celebrated peeling property, which guarantees universal bounds on the large-distance falloff of the Newman-Penrose (NP) scalars \cite{Penrose:1965am, Penrose:1972ea}. 
They are null projections of the Weyl tensor,
\begin{align}
\label{eq:NPscalars}
        \Psi_{4}  &=C_{\mu\nu\rho\sigma}\, N^{\mu}\varepsilon_{-}^{\nu}N^{\rho}\varepsilon_{-}^{\sigma} \ , 
        \quad
        \Psi_{3}  =C_{\mu\nu\rho\sigma}\, N^{\mu}L^{\nu}N^{\rho}\varepsilon_{-}^{\sigma}   \ ,                 
        \quad
        \Psi_{2}  =C_{\mu\nu\rho\sigma}\, N^{\mu}\varepsilon_{-}^{\nu}\varepsilon_{+}^{\rho}L^{\sigma}  \ ,      \\
        &\qquad\qquad\qquad
        \Psi_{1}  =C_{\mu\nu\rho\sigma}\, L^{\mu}N^{\nu}L^{\rho}\varepsilon_{+}^{\sigma} \ ,                   
        \qquad
        \Psi_{0}  =C_{\mu\nu\rho\sigma}\, L^{\mu}\varepsilon_{+}^{\nu}L^{\rho}\varepsilon_{+}^{\sigma} \ ,
        \nonumber
\end{align}
defined in terms of four null vectors (referred to as a \emph{null tetrad}) obeying\footnote{The index symmetrization is defined as $(ab)=(ab+ba)/2$.}
\begin{equation}
\label{eq:tetrad_props}
    N \cdot L = 1\ , \quad \varepsilon_{+} \cdot \varepsilon_{-} = -1
    \ , \quad 
    \varepsilon_{+} \cdot N =  \varepsilon_{+} \cdot L = 0 
    \ , \quad
     g^{\mu \nu} = 2 N^{(\mu} L^{\nu)} - 2 \varepsilon_{+}^{(\mu} \varepsilon_{-}^{\nu)} \ . 
\end{equation}
The peeling theorem, which was initially derived under stronger assumptions than asymptotic simplicity \cite{Sachs:1961zz, Penrose:1962ij}, implies that\footnote{Peeling conditions are distinct from a set of weaker conditions identified by Christodoulou in Ref.~\cite{christodoulou1993global}. The latter conditions, which sufficient to prove the global stability of asymptotically-Minkowski spacetimes:
\begin{equation*}
\lim_{\r\rightarrow \infty}\Psi_{1},\Psi_{0}\sim 
\mathcal{O}\left(\r^{-7/2}\right) \ .
\end{equation*}
}
\begin{equation}\label{eq:sachssmooth}
\lim_{\r\rightarrow \infty}\Psi_{k}(x^{\mu}) \sim \mathcal{O}(\r^{k-5}) \ ,
\end{equation}
where $\r$ is the affine parameter along a null geodesics~\cite{Sachs:1961zz,Penrose:1972ea} and the limit is taken towards future null infinity. A failure of the peeling property serves as a clear diagnostic for the breakdown of asymptotic simplicity. 
Since we are interested in identifying the leading terms in the asymptotic expansion of the components of the Weyl tensor, we will only consider the tetrad to leading order in $\r^{-1}$, \textit{i.e.} at zeroth order. We will choose 
\begin{equation}
\label{eq:metric_tetrad}
    \begin{split}
        L^\mu                = \frac{1}{\sqrt{2}}(t^\mu + r^\mu)\ , \qquad
        N^\mu                = \frac{1}{\sqrt{2}}(t^\mu - r^\mu)\ , \qquad
        \varepsilon_{-}^\mu  = (\varepsilon_{+}^\mu)^*\ ,
    \end{split}
\end{equation}
where
\begin{equation}\label{posdef}
    x^\mu = t \,  t^\mu + \r r^\mu\ , \quad \text{with } t^{2}=1\,,\; r^2 = -1\,,\;\text{and } t\cdot r = 0\,,
\end{equation}
is the observer's location, i.e. the location where we measure the NP scalars. 

Investigations~\cite{Damour:1985cm,Christodoulou:2002,Winicour:1985} have shown that generic scattering processes of massive particles do not exhibit peeling in the low-velocity limit and therefore also do not exhibit asymptotic simplicity,\footnote{While a scattering configuration may not appear isolated and thus may appear as departing from the initial intuitive argument, distant observers will be able to observe such a system of massive particles and experience its gravitational radiation while also seeing it as isolated. In other words, there is a clear hierarchy between the characteristic scale of the scattering and the observer’s distance from the event.} casting doubt on the conjecture that all physically-relevant spacetimes should be asymptotically simple. Even seemingly benign initial data can evolve into configurations where the expected fall-off of NP scalars~\eqref{eq:sachssmooth} at ${\cal I}^+$ breaks down~\cite{Kehrberger:2021uvf}.
These deviations arise because the dynamical components of the waveform, governed by the soft theorems for generic gravitational scattering processes \cite{Das:2023onv,Choi:2024ajz}, fail to decay sufficiently rapidly at large distances and late times \cite{Damour:1985cm}. 
Given this known departure from asymptotic simplicity,\footnote{Violations of peeling due to logarithmic terms in the soft expansion of the frequency-space scattering waveform, and their implications for the definition of asymptotic symmetries have been examined in Refs.~\cite{Geiller:2022vto,Geiller:2024ryw,Fuentealba:2024lll}. Although some scenarios remain unstudied, current results indicate that asymptotic symmetries can still be defined even in the absence of the peeling property~\eqref{eq:sachssmooth}.} it is natural to inquire about the violation of asymptotic simplicity in generic physical processes.
To address this, we study fully-relativistic scattering events as prototypical examples of physical gravitational processes.\footnote{The study of the peeling property for bound two-body systems whose metric admits a Multipolar post-Minkowskian expansion and is stationary in the far past is considered in Refs.~\cite{Walker:1979zk,Blanchet:1985sp, Blanchet:1986dk}.}
We focus on whether peeling fails at future, rather than past, null infinity, which we view as more physically relevant. Specifically, we explore whether generic, physically reasonable Cauchy data evolve to spacetimes that do not exhibit peeling at future null infinity.

The recent detection of gravitational waves by the LIGO and Virgo collaborations~\cite{LIGOScientific:2016aoc,LIGOScientific:2017vwq} 
has catalyzed the development of novel quantum field theory approaches 
to the classical relativistic two-body problem in general relativity.
Significant advances have been made in computing classical scattering observables, such as the scattering angle \cite{Bern:2021yeh,Bern:2021dqo,Dlapa:2021npj,Dlapa:2021vgp,Dlapa:2022lmu,Jakobsen:2023ndj,Jakobsen:2023hig, Damgaard:2023ttc} and gravitational waveform \cite{Cristofoli:2021vyo,Jakobsen:2021smu,Mougiakakos:2022sic,Riva:2022fru,Herderschee:2023fxh,Georgoudis:2023lgf,Brandhuber:2023hhy,Caron-Huot:2023vxl,Bini:2024rsy,Georgoudis:2024pdz, Bohnenblust:2025gir, Bohnenblust:2023qmy}, i.e. the leading $1/\r$ order in the large-$\r$ expansion of the spacetime metric sourced by a scattering process.
These methods also shed light on other aspects of general relativity, such as gravitational memory~\cite{Strominger:2014pwa} and the radiation reaction and the tail effect~\cite{Bern:2021yeh, Dlapa:2021npj, Dlapa:2021vgp, Jakobsen:2023hig, Jakobsen:2023ndj, Jakobsen:2022psy,Kalin:2022hph, Elkhidir:2023dco,Driesse:2024xad,Driesse:2024feo}, and sparked a renewed understanding of the universal properties of soft gravitational radiation~\cite{Saha:2019tub, Sahoo:2021ctw, Alessio:2024onn}. 
They were also used to (re)construct static metrics throughout space-time, such as the Schwarzschild solution, both perturbatively around the origin and to all orders, see e.g. Refs.~\cite{Boulware:1968zz,Duff:1973zz,Bjerrum-Bohr:2002fji,Neill:2013wsa,Jakobsen:2020ksu,Mougiakakos:2020laz,DOnofrio:2022cvn,Mougiakakos:2024nku,Monteiro:2014cda,Mougiakakos:2024nku,Damgaard:2024fqj,Mougiakakos:2024lif}. 

In this paper, we develop a framework to explore the asymptotic simplicity of spacetimes sourced by relativistic scattering systems by studying their peeling properties.
We define the departure of the spacetime metric from Minkowski space away from future null infinity, with complete velocity dependence, as a suitable expectation value of the graviton field and compute it using an off-shell generalization of the observable-based (KMOC) formalism~\cite{Kosower:2018adc, Cristofoli:2021vyo}. 
Since the spacetime metric produced by such calculations is typically not in Bondi coordinates, so we will directly use the linearized NP scalars as probes of peeling properties.
The null projections of the nonlinear terms are found using the factorization properties of correlation functions in the classical limit~\cite{Cristofoli:2021jas}.
We illustrate this framework by applying it to the leading order and the next-to-leading order NP scalars for the spacetime metric sourced by a two-body scattering in the post-Minkowskian expansion.

While the methods mentioned above easily generalize to compute the Fourier transform of the metric, which we sometimes loosely refer to as `the metric in momentum space', the required (inverse) Fourier transform to position space exhibits some interesting structure in the large-$\r$ expansion, related in part to the off-shellness of the outgoing gravitons. 
Using the method of regions~\cite{Beneke:1997zp}, we identify two momentum configurations for the external graviton momentum, which give distinct contributions to the Fourier transform. We refer to as the \textit{radiation} and \textit{Coulomb} regions, respectively. They both correspond to almost on-shell momenta;
in the former, the space-like outgoing graviton momentum is $k^\mu\sim {\cal O}(\r^0)$, while in the latter it is $k^\mu\sim {\cal O}(\r^{-1})$ at large $\r$.
We stress that these regions refer {\em only} to the Fourier transform of the external graviton momentum, which is already soft from the perspective of the scattering process. 

At tree level, we show that the radiation region exhibits the expected peeling behavior, but the $\mathcal{O}(G^{2})$ Coulomb region does not, yielding
\begin{equation}
\label{eq:Christodouloucond}
 \Psi_{0} \sim \frac{1}{\r^{4}} \ ,\qquad \Psi_{1}\sim \frac{\log \r}{\r^{4}}\ .
\end{equation}
This contribution corresponds to the breakdown of asymptotic simplicity found by Damour and Christodoulou in~Refs.~\cite{Damour:1985cm,Christodoulou:2002}.\footnote{Although Ref.~\cite{Christodoulou:2002} claims that the two results to be equivalent, this connection is not obvious. We thank Thibault Damour for extensive discussions on this point.}

The Fourier transform of the higher-loop momentum-space metric is substantially more involved. We set up a general framework to fully carry out such calculations and apply it at one-loop order, targeting the specific terms that lead to peeling violation. 
At one-loop, the radiation region contains a novel source of breaking of the peeling property, which is more severe than that of the Coulomb region, coming from tail terms that encode the back-scattering of gravitational waves on the spacetime curvature interaction.
This novel peeling violation is directly tied to the non-analytic behavior of the momentum-space metric as a function of $k^{2}$, encoding the infinite range of gravitational interactions and the (essentially) infinite (Shapiro) time delay the emitted graviton experiences as it climbs out of the potential well of the source.\footnote{We note the presence of an additional infrared divergence associated with the propagation of the massive bodies toward their point of closest approach, which is not tied to peeling violation in the far future.} Concretely, we find that $\mathcal{O}(G^{3})$ radiation region gives
\begin{equation}\label{onelooppeelvio}
\ \Psi_{0}\sim \frac{1}{\r^{4}}\ ,\qquad \Psi_{1}\sim \frac{1}{\r^{3}} \ .
\end{equation}
This is a much more drastic departure from the peeling property than at tree level,
which for $\Psi_1$ indicates a slower falloff than the one identified by Christodoulou and quoted in Eq.~\eqref{eq:Christodouloucond}.
We believe that such a falloff has not been previously observed in a 
scattering context. We leave for the future a thorough exploration of its consequences on asymptotic symmetries.

\paragraph{Outline of the paper.} The paper is organized as follows. In Sec.~\ref{sec:asymptotic_expansion}, we set up the general framework to compute the finite-distance metric from a QFT-based approach as the one-point function of the graviton field. In particular, we provide a generalization of the KMOC formalism to off-shell expectation values, an analyses of the relevant scales and an explicit map to obtain the needed information for the asymptotic analysis by recycling the on-shell computation. In Sec.~\ref{sec:asymptotic_analysis}, we describe the technical approach to the asymptotic expansion of the metric toward future null infinity. We identify two regions contributing to the Fourier transform from momentum to position space, in which the graviton momenta are $k^\mu \sim \r^{-1}$ and $k^\mu \sim \r^{0}$. We refer to them as the \textit{Coulomb and radiation regions}, respectively. The relevant computations are illustrated in Secs.~\ref{sec:radiationreg}~and~\ref{subsec:coulombregion}. Further technical details on a convenient gauge choice, and the construction of the Weyl tensor and the NP scalars are given in Sec.~\ref{metricansatz}. The main results of the paper are discussed in Sec.~\ref{sec:peelinganalysis}. First we discuss the Coulomb region contribution to the metric: in Sec.~\ref{leadingsoft} we study its leading contribution for generic retarded times, as well as $|u| \sim \r$, reproducing the \textit{memory} displacement between the waveform at early and late retarded times; while, in Sec.~\ref{sec:coulomb_region}, we focus on finite retarded times, $ |u| \ll \r$, and show that the subleading contribution in the Coulomb region leads to the violation of the peeling previously identified in Refs.~\cite{Damour:1985cm,Christodoulou:2002}. The analysis of the radiation region is split into two parts: in Sec.~\ref{sec:leadingwaveform} we focus contributions that are analytic at $k^2=0$, and in Sec.~\ref{sec:sublead} we analyze the non-analytic contributions, due to long-range non-linear interactions, \textit{i.e.} the \textit{tails}. In both cases, we analyze the contributions from the corrections to the tetrad, the non-linearities of the NP scalars in the metric, as well as from singularities at finite distance in the complex $k^\mu$ plane. In particular, in Sec.~\ref{analytic}, we generically proof any term which is analytic at $k^2=0$ must respect the peeling property. While, in Sec.~\ref{sec:1looplogs}, combining dimensional analysis with the angular dependence of the Weyl tensor in momentum space, we show that the tails lead to a more severe violation of the peeling property than previously recognized. Finally, we conclude in Sec.~\ref{sec:conclusions}, raising some open questions about the tension of our new results with the existing results in the literature, and proving a list of future directions.

\section{The finite distance metric and off-shell non-analytic terms}
\label{sec:asymptotic_expansion}

As discussed, our goal is to compute the NP scalars for the spacetime corresponding to a two-body scattering process and use them 
as a diagnostic of asymptotic simplicity. The NP scalars $\Psi_i$ are nonlinear functions of the metric. In the classical limit and assuming the point-particle approximation, the one-point function $\langle \Psi_i\rangle$ factorizes into a function of the one-point function of its constituents~\cite{Cristofoli:2021jas},
\begin{equation}
\langle \Psi_i(\hat{g}(x))\rangle|_{\hbar
\rightarrow 0} =  \Psi_i(\langle\hat{g}(x)\rangle|_{\hbar
\rightarrow 0} ) \ ,
\end{equation}
which follows from the suppression of quantum uncertainty in the classical limit and the deterministic nature of long-range interactions. For asymptotically Minkowski spaces, $g_{\mu\nu} = \eta_{\mu\nu} + h_{\mu\nu}$, so it suffices to study $h_{\mu\nu}$.\footnote{Note that we absorb the factor of $\kappa$ in the usual definition of the metric fluctuation, $g_{\mu\nu} = \eta_{\mu\nu} + \kappa h_{\mu\nu}$, into the fluctuation field. The leading order correction to the metric remains ${\cal O}(G^2)$.} Thus, we will focus on $h_{\mu\nu}$ for the rest of this section. Linear operators acting on $h_{\mu\nu}$ can trivially be included. We will not spell the precise procedure out here, but will use it in later sections to compute, e.g., the linearized Weyl tensor or the linearized NP scalars.

\subsection{The finite distance metric from the final-state graviton one-point function}

In the presence of a source $J$ (here the two-body scattering process), the spacetime metric is given by the graviton 1-point function, schematically,
\begin{align}
\label{eq:1ptfct}
h_{\mu\nu}(x) = \langle {\hat h}_{\mu\nu}(x) \rangle_J \ .
\end{align}
The leading ${\cal O}(\r^{-1})$ part of this metric is invariant under gauge transformations that fall off sufficiently fast.
The original formulation of the observables-based (KMOC) formalism~\cite{Kosower:2018adc,Cristofoli:2021vyo} computes 
the final-state one-point function of the linearized Newman-Penrose scalar $\Psi_4$, in terms of on-shell S-matrix elements, where the linearized Newman-Penrose scalar $\Psi_4$ is given by second time derivatives of $h_{\mu\nu}(x)$.
The same formalism was used in Refs.~\cite{Herderschee:2023fxh,Georgoudis:2023lgf,Brandhuber:2023hhy,Caron-Huot:2023vxl,Bini:2024rsy, Bohnenblust:2025gir, Bohnenblust:2023qmy} to compute directly the waveform as the one-point function of the graviton field.
The evaluation of higher-order contributions in the large-distance expansion requires keeping the graviton off-shell. 

Starting with a state $|\Psi_\text{in}\rangle$ describing two massive on-shell incoming particles at $t=-\infty$, the finite-time state needed for the evaluation of the 1-point function~\eqref{eq:1ptfct} is 
\begin{align}
    |\Psi,t\rangle = U(t, -\infty) |\Psi_\text{in}\rangle \ ,
\end{align}
where $U(a, b)$ is the time-evolution operator from $b$ to $a$. 
Then, the metric perturbation $h_{\mu\nu} = g_{\mu\nu} - \eta_{\mu\nu}$ at finite distance is given by the expectation value of the corresponding operator $\hat{h}_{\mu\nu}$ in this state 
\begin{equation}
    \begin{split}
    h_{\mu\nu}(x) & =
    \langle \Psi,t | {\hat h}_{\mu\nu}(x) |\Psi,t\rangle
    -
    \lim_{t\rightarrow -\infty}\langle \Psi ,t| {\hat h}_{\mu\nu}(x) |\Psi,t\rangle
    \\
                  & =
    \langle \Psi_\text{in} | U^\dagger(-\infty, t) {\hat h}_{\mu\nu}(x) U(t, -\infty) |\Psi_\text{in}\rangle
    \\
                  & -
    \lim_{t\rightarrow -\infty}\langle \Psi_\text{in} | U^\dagger(-\infty, t) {\hat h}_{\mu\nu}(x) U(t, -\infty) |\Psi_\text{in}\rangle
    \end{split}
\end{equation}

Then, we insert a conjugated final-state identity operator as
\begin{align}
    \id
    = U^\dagger(t, \infty) \, \id_\text{fin} \, U(+\infty, t)
    =\sum_\text{fin}  U^\dagger(t,\infty) |n_\text{fin}\rangle \langle n_\text{fin}|U(+\infty, t)
\end{align}
where the sum is over all final states. This leads to
\begin{align}
    \label{eq:h}
    h_{\mu\nu}(x) & = \sum_\text{fin}
    \langle \Psi_\text{in} | U^\dagger( -\infty, t) U^\dagger(t,+\infty) |n_\text{fin}\rangle \langle n_\text{fin}|U(+\infty, t) {\hat h}_{\mu\nu}(x)  U(t, -\infty) |\Psi_\text{in}\rangle
    \\
                  & -
    \lim_{t\rightarrow -\infty}\sum_\text{fin}
    \langle \Psi_\text{in} | U^\dagger(-\infty,t) U^\dagger(t,+\infty) |n_\text{fin}\rangle \langle n_\text{fin}|U(+\infty, t) {\hat h}_{\mu\nu}(x)  U(t, -\infty) |\Psi_\text{in}\rangle
    \nonumber
\end{align}
The two factors on the first line are anti-time-ordered and time-ordered, respectively. The first factor in the summand, $\langle \Psi_\text{in} | U^\dagger(-\infty, t) U^\dagger(t, \infty) |n_\text{fin}\rangle = \langle n_\text{fin} | U(+\infty, -\infty)| \Psi_\text{in} \rangle^*$, is the operator definition of the conjugated $S$-matrix element between the initial state $|\Psi_\text{in}\rangle$ and the final state $|n_\text{fin}\rangle$. The second factor, $\langle n_\text{fin}|U(+\infty, t) {\hat h}_{\mu\nu}(x)  U(t, -\infty) |\Psi_\text{in}\rangle$, is the operator definition of the 
form factor of the operator ${\hat h}_{\mu\nu}(x)$.
The leg corresponding to $\hat{h}_{\mu\nu}$ in the first term is not amputated and has a retarded $i\epsilon$ prescription. Importantly, the second term is not identically zero; rather, it subtracts the constant contribution coming from the metric of the two objects when they are very far from each other.\footnote{The graviton one-point function can also be defined in the worldline formalism with the retarded propagators of Refs.~\cite{Kalin:2020mvi,Mogull:2020sak,Jakobsen:2021smu}. It would be interesting to explore the equivalence of these definitions, perhaps along the lines of Ref.~\cite{Capatti:2024bid}.}

\begin{figure}[t]
    \centering
    \begin{tikzpicture}
        \pgfmathsetmacro{\w}{5.5}
        \pgfmathsetmacro{\h}{1.5}
        \node [left=0pt] (p1) at (-\w,0) {$1$};
        \node [right=0pt] (p4) at (\w,0) {$1$};
        \node [left=0pt] (p2) at (-\w,\h) {$2$};
        \node [right=0pt] (p3) at (\w,\h) {$2$};
        \node [left=0pt] (p5) at (\w/4,{3*\h/2}) {$h^{\mu\nu}(x)$};
        \draw [very thick] (p1) -- (p4);
        \draw [very thick,blue] (p2) -- (p3);
        \draw [graviton] (-\w/2,\h/2) -- (\w/2,\h/2) (-\w/2,\h/2+0.3) -- (\w/2,\h/2+0.3) (-\w/2,\h/2-0.3) -- (\w/2,\h/2-0.3);
        \draw [graviton,double,thick] (\w/2,\h) to[bend right=45] (p5.east);
        \filldraw [fill=gray!30!white,thick] (-\w/2,\h/2) ellipse (0.25cm and 1cm);
        \filldraw [fill=gray!30!white,thick] (\w/2,\h/2) ellipse (0.25cm and 1cm);
        \draw [line width=3pt,draw=white] (0,{3*\h/2}) -- (0,-\h/2);
        \draw [dashed,draw=red] (0,{3*\h/2}) -- (0,-\h/2);
        \node [below=0.25cm] at (-\w/2,0) {\scalebox{0.8}{$\langle\Psi_{\text{in}}|U^\dagger(-\infty,t)U^\dagger(t,+\infty)\vphantom{\hat{h}_{\mu\nu}}|n_{\text{fin}}\rangle$}};
        \node [below=0.25cm] at (\w/2,0) {\scalebox{0.8}{$\langle n_{\text{fin}}|U(+\infty,t)\hat{h}_{\mu\nu}(x)U(t,-\infty)|\Psi_{\text{in}}\rangle\vphantom{U(+\infty,t)^{\dagger}}$}};
    \end{tikzpicture}
    \caption{Typical diagrams contributing to the final-state graviton one-point function in Eq.~\eqref{eq:h}.}
    \label{fig:typicall_term}
\end{figure}
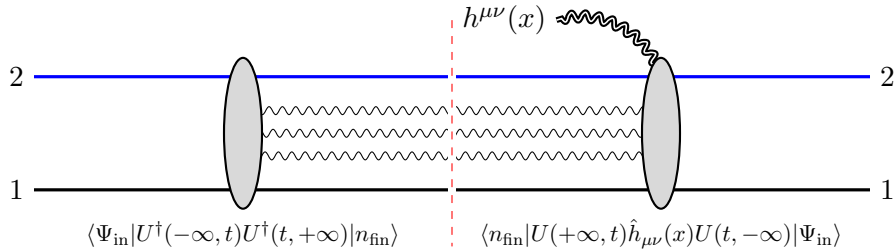

The first line of Eq.~\eqref{eq:h} is the direct generalization of the $\langle \textrm{in}|\hat{S}^{\dagger}\hat{h}_{\mu\nu}\hat{S}|\textrm{in}\rangle$ term in the original waveform calculation \cite{Cristofoli:2021vyo}, see Fig.~\ref{fig:typicall_term}. 
As in that case, the factors
\begin{equation}
    \langle \Psi_\text{in} | U^\dagger(-\infty,t) U^\dagger(t, +\infty ) |n_\text{fin}\rangle,\quad \textrm{and} \quad  \langle n_\text{fin}|U(+\infty, t) {\hat h}_{\mu\nu}(x)  U(t, -\infty) |\Psi_\text{in}\rangle
\end{equation}
contain both connected and disconnected graphs. 
In the classical limit, the intermediate states $|n_\text{fin}
\rangle$ contain the two massive particles and any number of massless
particles. In the sum over states, this formula exhibits the same cancellations that are familiar from the waveform, see e.g.~\cite{Cristofoli:2021vyo,Brandhuber:2023hhy,Herderschee:2023fxh,Caron-Huot:2023vxl}.
For example, all the classically singular contributions cancel. 

Similar to Ref.~\cite{Kosower:2018adc, Cristofoli:2021vyo}, our initial state is a two-particle state:
\begin{equation}\label{psiin}
    |\Psi_\text{in} \rangle =\int\! \frac{d^{d}p_{1}}{(2\pi)^{d}}\frac{d^{d}p_{2}}{(2\pi)^{d}}\hat{\delta}^{+}(p_{1}^{2}-m_{1}^{2})\hat{\delta}^{+}(p_{2}^{2}-m_{2}^{2})e^{ip_{1}\cdot b_{1}+ip_{2}\cdot b_{2}}\phi(p_{1})\phi(p_{2})|p_{1},p_{2}\rangle
\end{equation}
where
\begin{equation}
    \hat{\delta}^{+}(p_{i}^{2}-m_{i}^{2})\equiv2\pi \delta(p_{i}^{2}-m_{i}^{2})\Theta(p^{0}_{i}) 
\end{equation}
and $|p_{1},p_{2}\rangle$ is a two-particle momentum eigenstate. The classical wavefunctions are sharply peaked around some (classical) value of the incoming momenta and are normalized to unity
\begin{equation}
    1=\int\! \frac{d^{d}p_{i}}{(2\pi)^{d}}\hat{\delta}^{+}(p_{i}^{2}-m_{i}^{2}) |\phi(p_{i})|^{2} \ .
\end{equation}
We substitute Eq. (\ref{psiin}) into Eq. (\ref{eq:h}) and consider the classical limit of the expression. 

The construction we presented works in the quantum theory.
A careful choice of wavepackets corresponds to taking the classical limit, as outlined in the KMOC formalism. In particular, the wavepackets have to be such that the variation of \textit{any} observable for such a state is much smaller than the (squared) expectation value. This is equivalent to the classical limit in the sense of Bohr's Correspondence Principle, which equates the classical limit with the limit of large changes -- \textit{i.e.} large mass and angular momenta:
\begin{equation}
    \frac{1}{m_i} \ll R_i\ ,\quad \frac{1}{m_i} \ll b\ ,\quad \frac{1}{m_i v_i} \ll b\ ,
\end{equation}
where $m_i$ is the mass of the scattering objects, $v_i$ their velocities, $R_i$ is their characteristic size (\textit{e.g.} for black holes the \textit{only} characteristic length is their Schwarzschild radii $R_i=G m_i$) and $b$ is the impact parameter. Moreover, we will restrict to the weakly-interacting set-ups,
\begin{equation}
    R_i \ll b \, v^2 < b\ ,
\end{equation} 
($b v^2$ is a measure of the minimum approach distance)
which will allow us to use perturbation theory. In particular, this hierarchy of scales corresponds to the so-called \textit{post-Minkowskian} approximation. Finally, we will consider the expectation value of the metric in the far future at a distance $r=\r \sim t \, $ which is larger than any other typical length of the system:
\begin{equation}
    \frac{1}{m_i} \ll R_i \ll b \ll r\ ,
\end{equation}
and perform an asymptotic expansion at large values of $r$.

We assume that the incoming wavepackets are sharply peaked Gaussian wavefunctions.
The end result of this calculation is an integral of the form
\begin{equation}\label{eq:momentumspaceexpress}
\begin{split}
    h^{\mu\nu}(x)&=-\int\! \frac{d^{d}k}{(2\pi)^{d}} \frac{e^{-i k\cdot x}}{(k_{0}+i\epsilon)^2-\vec{k}^{2}}\mathcal{T}^{\mu\nu}_{\mu'\nu'}J^{\mu'\nu'}(k) 
    \equiv -\int\! \frac{d^{d}k}{(2\pi)^{d}} \frac{e^{-i k\cdot x}\htilde^{\mu\nu}(k)}{(k_{0}+i\epsilon)^2-\vec{k}^{2}}
    \end{split}
\end{equation}
where\footnote{In this paper, we find it convenient to work in an axial gauge, as it does not involve spurious extra $\frac{1}{k^2}$ factors, like in the de~Donder gauge.}
\begin{equation}
\begin{split}
    \label{eq:projector}
    \mathcal{T}^{\mu\nu}_{\mu'\nu'}&=\frac{1}{2}\left ( \mathcal{P}_{\mu'}^{\mu}\mathcal{P}_{\nu'}^{\nu}+\mathcal{P}_{\nu'}^{\mu}\mathcal{P}_{\mu'}^{\nu}-\frac{2}{d-2}\mathcal{P}^{\mu\nu}\mathcal{P}_{\mu'\nu'}\right )\ , \\
    \mathcal{P}^{\mu\nu}&=\eta^{\mu\nu}-\frac{\zeta^{\mu}k^{\nu}+\zeta^{\nu}k^{\mu}}{\zeta\cdot k} + \zeta^2 \frac{k^\mu k^\nu}{(\zeta \cdot k)^2}\ .
    \end{split}
\end{equation}
$\mathcal{T}$ in Eqs.~\eqref{eq:momentumspaceexpress} and \eqref{eq:projector} is the numerator of the graviton propagator, $\zeta$ is an arbitrary (reference) vector and $J^{\mu\nu}(k)$ denotes the classical source generated by the scattering, including the matter stress-tensor and the gravitational non-linearities. $\htilde^{\mu\nu}(k)$ is not the direct Fourier transform of the metric due to the presence of the retarded propagator.
While this is a non-standard numerator for the graviton propagator, it has the feature that it is proportional to $k^2$ when either traced or contracted with the graviton momentum. This feature will be useful in Sec.~\ref{sec:sublead} when identifying the leading non-analyticities as $k^2\rightarrow 0$.

Interpreting Eq.~\eqref{eq:momentumspaceexpress} as the solution to the linearized Einstein's equation, gauge invariance requires that the source $J^{\mu\nu}(k)$ be transverse,
\begin{equation}
k_\mu J^{\mu\nu}(k) = 0 \ ,
\end{equation}
where here $J^{\mu\nu}(k)$ stands both for the impact-parameter space source in Eq.~\eqref{eq:momentumspaceexpress} or for its momentum-space version. Thus, given some off-shell source, we must gauge-transform it so that it has this property.

\subsection{Off-shell and on-shell non-analyticity of the final-state graviton one-point function}
\label{sec:offshellgravitonregularization}

Scattering amplitudes are typically infrared divergent even in the classical limit. Dimensional regularization, defining then in $d=4-2\epsilon$ dimensions, provides a gauge-invariant regularization scheme. Since we are interested in computing the metric at finite distance, the momentum associated to the graviton is to be considered off-shell, \textit{i.e.} $k^2\neq 0$. While this off-shellness regularizes the infrared divergences associated with the outgoing external graviton, it is not a substitute of dimensional regularization which is still required to regularize certain certain initial-state singularities. In this section we describe a systematic procedure to obtain the leading logarithms in $k^2$ of the relevant integrals from the dimensionally-regulated ones.

The LSZ reduction guarantees that the graviton one-point function differs from its on-shell version by terms that vanish on-shell (i.e. that are proportional to $k^2$ and $\varepsilon_{\mu\nu}(k) k^\mu$) and by terms connected to the relation between the on-shell and off-shell regulators. 
In later sections, we will demonstrate in examples and argue in general that additional factors of $k^2$ and $\varepsilon_{\mu\nu}(k) k^\mu$ compared to the on-shell amplitude lead only to subleading terms in the asymptotic expansion of the position-space scattering metric and that such terms exhibit Sachs' peeling property. 
Thus, with the benefit of hindsight, we will ignore these terms and focus on the regularization-related ones. It is moreover natural to expect that the terms that are non-analytic in the components of the graviton momentum behave differently under the Fourier transform~\eqref{eq:momentumspaceexpress} than the analytic ones; we will therefore separate the former below.

An off-shell graviton momentum, i.e., $k^2\ne 0$,  regularizes the IR divergences of correlation functions from diagrams with three-graviton vertices. These are the integrals that contribute \textit{e.g.} to the ${\cal O}(G^3)$ IR divergences of Ref.~\cite{Weinberg:1965nx} and to virtual IR divergences at all loop orders.
Thus, on dimensional grounds, we may expect that, on the one hand, there exist terms that are nonanalytic in $k^2$ and, on the other, that they can be obtained via a simple map between the dimensional and off-shell regulators. 

To demonstrate this idea and also establish the regulator map, we recompute the box integral with one three-graviton vertex, denoted as $I_{1,1,1,0}$ in Ref.~\cite{Herderschee:2023fxh}, while keeping $k^2\ne 0$. The result, expanded at small $k^2$, is 
\begin{align}
    I_{1,1,1,0} &= \int\frac{\dd^4\ell}{(2\pi)^4}\frac{\hdelta(2u_2\cdot\ell)}{\ell^2(\ell+q_2)^2(\ell-q_1)^2} \nonumber\\
    &= \frac{1}{16 q_2^2 (u_2 \cdot k)} + \frac{i}{16 \pi q_2^2 (u_2\cdot k)} \log\frac{4(u_2\cdot k)^2 q_2^2}{(-k^2)q_1^2} + \mathcal{O}\left(\frac{k^2}{(u_2\cdot k)^2}\right)\,.
\label{eq:offshell_regularization}    
\end{align}
where we have assumed that $k^2<0$.
The numerator under the logarithm, $(u_2\cdot k)^2$, is also present in the on-shell calculation, and captures the interaction of the outgoing graviton with its matter source.
The same integral has been computed with dimensional regularization and $k^2=0$ for the computation of the one-loop waveform, see \textit{e.g.} Eq.~(B.5) of Ref.~\cite{Herderschee:2023fxh}.
Comparing with eq.~\eqref{eq:offshell_regularization}, we see that the regulator dependent part are related, as expected, 
by the simple replacement
\begin{align}
\label{replacement}
\frac{1}{\epsilon} -\gamma_E+\log(4\pi^2)  \equiv  \frac{1}{\epsilon_\text{IR}} ~\longmapsto~ 
 2\log \frac{\pi(-k^2)}{\mu^2} \ ,
\end{align}
where $\mu$ is the dimensional regularization scale.\footnote{The real part of eq.~\eqref{eq:offshell_regularization} turns out to differ from the same result in dimensional regularization computed in Refs.~\cite{Brandhuber:2023hhy,Herderschee:2023fxh} (the rational part differ by a factor of two, as well as the relative coefficient between the different logarithms).
Despite this apparent difference, we believe that the two regularizations will still result in the same waveform up to a different, and scheme-dependent, phase.
Indeed, we note that waveform result will look very different from the one computed within dimensional regularization and on shell: with the current $k^2 \neq 0$ regularization, the master integral basis for the waveform will be different due to this extra scale, and so will be their coefficients due to different IBP identities. We expect this to reconcile the difference in individual master integrals. An explicit confirmation is of course desirable, but it is beyond the scope of the present paper.
}
The same holds for the other integrals with a three-graviton vertex.
Thus, the ${\cal O}(G^3)$ metric can be written as
\begin{equation}
    \label{eq:infrared_divergences}
    \htilde^{\mu \nu}_{\text{1-loop}} = - 2i G (m_{1}\, u_{1} + m_{2} \, u_{2} ) \cdot k\,\log \left[\frac{\pi(-k^2)}{\mu^2}\right] 
    \htilde^{\mu \nu}_{\text{tree}} + \text{(analytic in $k^2$)} \ .
\end{equation}
The first term is equivalent to Weinberg's classic result~\cite{Weinberg:1965nx} up to a map on the IR regulators, with the exception that, while $h^{\mu \nu}_{\text{tree}}$ has the same functional form as the tree-level amplitude with stripped polarization tensor, the graviton momentum is taken off-shell.
The second term, $\widetilde h^{\mu \nu}_{\text{1-loop}}$, 
contains all the analytic terms in $k^2$. 
We will show in Sec.~\ref{sec:peelinganalysis} that these terms do contribute to the NP scalars in accordance with Sachs's peeling property, so we will not consider them explicitly.

The replacement~\eqref{replacement} effectively sets the dimensional regularization scale to be the scale of the outgoing graviton momentum. Viewed in position space, the regularization scale is determined by the distance between the observer and the scattering event.

We note that the off-shellness of the graviton momentum does not regularize all infrared divergences, so a dimensional regulator is still necessary to evaluate the final-state graviton one-point function. 
Indeed, the propagators on the matter lines appear with a retarded $i\epsilon$ prescription. While the time-symmetric part of these contributions does not need extra regulators in the off-shell one-point function, the delta function introduces extra IR divergences associated with the incoming matter states~\cite{Caron-Huot:2023vxl}.
While these divergences can be regularized by taking the matter particles off shell (or, equivalently, preparing the initial state at a finite distance), this will effectively replace the dimensional regularization scale of these divergences by the virtuality of the matter particles. Such terms will not affect the Fourier transform to position space, Eq.~\eqref{eq:momentumspaceexpress}, so we will not consider this explicitly here. In particular, the surviving $1/\epsilon_{\rm IR}$ divergence originates from the logarithmic drift of the incoming particles due to the long-distance nature of gravitational interactions.\footnote{This divergence can be removed by shifting the time coordinate or by keeping the incoming particles off-shell as previously mentioned. In the latter case, the external propagators must have advanced $i\epsilon$ prescriptions, as they correspond to the boundary conditions in the far past.}

Thus, the general form of the ${\cal O}(G^3)$ metric in momentum space, separated into nonanalytic and analytic parts, is
\begin{eqnarray}
    \label{eq:infrared_divergences_all}
    \htilde^{\mu \nu}_{\text{1-loop}} &=&  i G ( m_1 u_1 + m_2 u_2)\cdot k \left\{ 2\log\left[\frac{(u_1\cdot k)(u_2\cdot k)}{-k^2}\right] \right.
    \\
&&\left.   + \frac{\sigma(\sigma^2-3/2)}{(\sigma^2-1)^{3/2}}\left(-\frac{1}{\epsilon_{\rm IR}}+
    \log\left[\frac{(u_1\cdot k)^2(u_2\cdot k)^2}{\mu^4}\right]\right) \right\}
    \htilde^{\mu \nu}_{\text{tree}} + {\mathring{h}}^{\mu \nu}_{\text{1-loop}} \ ,
    \nonumber
\end{eqnarray}
where ${\mathring{h}}^{\mu \nu}_{\text{1-loop}}$ is analytic in all components of the graviton momentum and we will not need its explicit form.\footnote{The logarithmic branch cuts in Eq.~\eqref{eq:infrared_divergences_all} have specific $i\epsilon$ prescriptions, which follow from the explicit calculation of the final-state graviton one-point function. We will restore them in Sec.~\ref{sec:peelinganalysis} from consistency with the soft-graviton theorems and general properties of the position-space metric.}$^,$\footnote{In writing this expression, we chose a particular organization of the tail terms. In particular, ${\mathring{h}}^{\mu \nu}$ contains $(p_1-p_2)\cdot k \log (u_1\cdot k/u_2\cdot k) \htilde^{\mu \nu}_{\text{tree}}$.} Indeed, in Sec.~\ref{analytic}, we will present a proof that analyticity at $k^2=0$ implies the peeling conditions.

It is not difficult to construct an off-shell regularization of the IR divergences at higher-loop orders, and we will do so in Sec.~\ref{sec:exponentiated_FT}. The higher-loop analog of the second term in Eq.~\eqref{eq:infrared_divergences} requires a nontrivial calculation which we leave for future work.

\section{The asymptotic expansion of expectation values}
\label{sec:asymptotic_analysis}

While Eq.~\eqref{eq:momentumspaceexpress} defines the metric sourced by a scattering process at all space-time points, carrying out the integrals can be difficult. For our purpose, however, such a calculation
is unnecessary.
Indeed, a large-$\r$ asymptotic expansion of this metric should be sufficient to probe its peeling properties.
Furthermore, it is more convenient to evaluate Newman-Penrose scalars to probe gauge-invariant properties of spacetime.
In this section, we begin by discussing the asymptotic expansion of 
Eq.~\eqref{eq:momentumspaceexpress} large values of $\r$. Then, we upgrade the analysis to cover the evaluation of the asymptotic expansion of derivative operators acting on the metric and, in particular, of the Newman-Penrose scalars.

To this end,  we first parametrize the off-shell graviton momentum as
\begin{equation}
\label{eq:decomposition1}
    k^{\mu}= \omega\, t^\mu + \kk\, n^\mu\ ,\quad \text{where} \ n^{2}=-1 \text{ and } t\cdot n = 0
\end{equation}
and $t^{\mu}$ is given in Eq. (\ref{posdef}), such that
\begin{equation}
\label{eq:massageFT}
    \begin{split}
        \int\! \frac{d^{d}k}{(2\pi)^{d}} \frac{e^{-i k\cdot x} \, \htilde^{\mu\nu}(k)}{(k_{0}+i\epsilon)^2-\vec{k}^{2}} =\int_{-\infty}^{+\infty}\!\frac{d\omega}{2\pi}\int_{0}^{\infty}\! \frac{d\kk}{2\pi}\frac{e^{-i\omega t \cdot x} \kk^{d-2}}{(\omega+i\epsilon)^{2}-\kk^{2}}  \int\! dn \, e^{- i \kk n \cdot x}\htilde^{\mu\nu}(k) \ ,
    \end{split}
\end{equation}
where $dn$ is the measure on the $(d-2)$-dimensional sphere parametrized by the unit space-like vector $n^\mu$, normalized by the appropriate power of $2\pi$. 

The contour for the $\omega$ integral, as defined by the $i\epsilon$ prescription in Eq.~\eqref{eq:momentumspaceexpress}, encodes our desired (retarded) boundary conditions. The integration can be carried out via a contour deformation. In particular, since we are mainly focused on contributions in the far future, by deforming the $\omega$ integration contour to the lower-half plane, we pick up contributions from the singularities of the source and the external (retarded) propagator. This integral can typically be performed exactly.

The angular integration over $n$ can be further decomposed, using a position-space space-like vector $r^\mu$:
\begin{equation}
    x^\mu = t \,  t^\mu + \r r^\mu\ , \quad \text{where}\quad r^2 = -1\ .
\end{equation}
This additional direction provides a natural parametrization for the graviton momentum:
\begin{equation}
    \label{eq:parameterizationk}
    k^\mu = \omega t^\mu + \kk z r^\mu + \kk \sqrt{1-z^2} n_\perp^\mu\ , \quad \text{where}\quad n_\perp^2 = -1\, ,\ n_\perp \cdot t = 0\ \text{and}\ n_\perp \cdot r = 0\ ,
\end{equation}
where $z$ parametrizes the angle between the observed direction $\vec x$ and the direction $n$ of the outgoing graviton.  
The $n$ integral in Eq.~\eqref{eq:massageFT} then becomes
\begin{equation}
\label{eq:angular_integral}
\int\! dn\, e^{- i \kk n \cdot x}\,\htilde^{\mu\nu}(k) = 
    \int dn_{\perp} \int_{-1}^{1} \frac{dz}{2\pi} \,e^{i\kk z \r}\, (1-z^2)^{\frac{d-4}{2}} \, \htilde^{\mu\nu}(k) \ ,
\end{equation}
where $k^{\mu}$ in the argument of the current is given by Eq.~\eqref{eq:parameterizationk}, and $dn_{\perp}$ is the measure on the corresponding $(d-3)$-dimensional sphere. 

The $\kk$ integration is more subtle. At large distances, the integrand depends on two parametrically-separated scales: that of the system, $R_i \lesssim b$, and that of the observer, $\r$. Depending on the scaling of $\kk$ with the observer scale, the contribution of the oscillatory factor is different. Thus, we have two distinct integration regions, 
\begin{equation}
\label{eq:soft_hard_split}
\int_{0}^{\infty} d \kk\; \bullet=\underbrace{\int_{0}^{\lambda} d \kk \; \bullet}_{\textrm{Coulomb}}+\underbrace{\int_{\lambda}^{\infty} d \kk \; \bullet}_{\textrm{Radiation}} 
\qquad
\text{with}
\quad
R_i, b \ll \frac{1}{\lambda} \ll\r, \, t \ .
\end{equation}
Since the complete $\kk$ integral is convergent both at small $\kk$ (because of the known soft behavior of $\htilde^{\mu\nu}(k)$) and at large $\kk$ (because of the exponential suppression ensuing from the typical frequency of $\htilde^{\mu\nu}(k)$ being ${\cal O}(1/b)$), a divergence in one of the two regions should be regularized by the cutoff $\lambda$ and must cancel against a similar divergence in the other region.
Both of these regions contribute to the asymptotic expansion at large $\r$ and need to be considered separately. From the perspective of the method of regions in dimensional regularization~\cite{Beneke:1997zp,Jantzen:2011nz} applied on the LHS of \eqref{eq:massageFT},  the graviton momentum scales at large~$\r$ in these regions are 
\begin{align}
\label{eq:k_regions}
\text{Radiation region}: k^{\mu} \sim \r^0\ ,
\qquad\qquad
\text{Coulomb region}: k^{\mu} \sim \r^{-1}\ .
\end{align}
In the following, we work in dimensional regularization, dropping the regulator and setting $d=4$ whenever integrals are convergent. We will discuss the two regions in turn. 
It will turn out that the external gravitons are effectively close to on-shell in both regions. However, the space-like components in the Coulomb region are parametrically smaller than those in the radiation region.\footnote{\label{BMSfootnote}
Coulomb modes describe essentially static space-time metrics. They, however, also source nontrivial time-dependent contributions which can be interpreted as a certain BMS supertranslation 
of a metric with no Coulomb modes at ${\cal I}^+$, see e.g. Refs.~\cite{Veneziano:2022zwh, DiVecchia:2022owy, Georgoudis:2023eke, Bini:2024rsy, Elkhidir:2024izo}.
We also note that these time-dependent contributions bring logarithmic correction in the presence of off-shell external gravitons, somewhat analogously to the mechanism outlined in Sec.~\ref{sec:offshellgravitonregularization}, see \cite{Elkhidir:2024izo}.
In Sec.~\ref{sec:sublead}, we will discuss simultaneously all the $\log k^2$ terms, without distinguishing their detailed origin. Generically, different logarithmic correction correspond to different boundary conditions on space-like Cauchy hypersurfaces.
}

\begin{figure}[t]
    \centering
    \begin{tikzpicture}[every node/.style={font=\footnotesize}]
        \draw [stealth-stealth] (-3.75,0) -- (3.75,0) node[right=0pt]{Re[$z$]};
        \draw [-stealth] (0,0) -- (0,3.75) node[above=0pt]{Im[$z$]};

        \filldraw (-2,0) circle (1pt) node[below=0pt]{$z=-1$};
        \filldraw (2,0) circle (1pt) node[below=0pt]{$z=1$};

        \filldraw (4,4) circle (0pt) node[left=0pt]{$z$-plane};

        \draw [-stealth,thick,color=blue] (-2,0) -- (0,0);
        \draw [thick,color=blue] (0,0) -- (2,0);

        \draw [thick,color=red] (2,0) -- (2,1.5);
        \draw [thick,stealth-,color=red] (2,1.5) -- (2,3);

        \draw [thick,-stealth,color=red] (-2,0) -- (-2,1.5);
        \draw [thick,color=red] (-2,1.5) -- (-2,3);

        \filldraw (0,0) circle (0pt) node[below=0pt,color=blue]{Original contour};
        \filldraw (2,1.5) circle (0pt) node[right=0pt,color=red]{Right contour};
        \filldraw (-2,1.5) circle (0pt) node[left=0pt,color=red]{Left contour};
    \end{tikzpicture}
    \caption{The contour deformation of the $z$-integral. The original contour, denoted in blue, is equal to the sum of the advanced and retarded contours, denoted in red. We focus on the advanced contour.  
    }
    \label{fig:deformedcontour}
\end{figure}
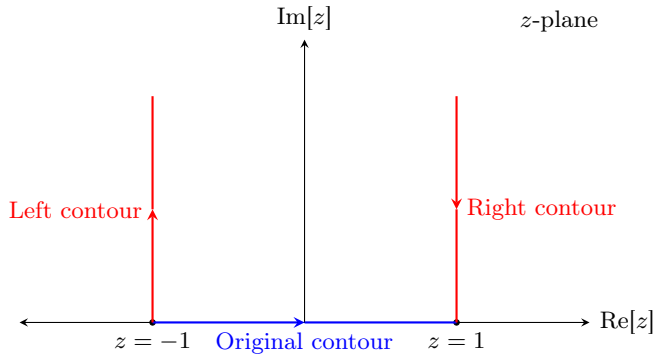

\subsection{The radiation region}\label{sec:radiationreg}

In this region, it is convenient to deform the $z$ contour into two contours parallel to the imaginary axis and connected at infinity, as in Fig.~\ref{fig:deformedcontour}. The integral takes the following form:
\begin{align}
    \label{eq:deformed_angular_contour_sec_3}
        &h^{\mu\nu}(x) =  \int_{-\infty}^{+\infty}\!\frac{d\omega}{2\pi i}\int_{0}^{\infty}\! \frac{d\kk}{2\pi}\frac{e^{-i\omega \, t-i \kk\,\r} \kk^{d-2}}{(\omega+i\epsilon)^{2}-\kk^{2}}\int\! dn_{\perp} \left [ \int_{0}^{\infty}\! \frac{dy}{2\pi} e^{- \kk y\r} (+2i y + y^2)^{\frac{d-4}{2}} \htilde^{\mu\nu}(k) \right ] \notag  \\
&\!\!
-\int_{-\infty}^{+\infty}\!\frac{d\omega}{2\pi i}\int_{0}^{\infty}\! \frac{d\kk}{2\pi}\frac{e^{-i\omega \, t+i\kk\r} \kk^{d-2}}{(\omega+i\epsilon)^{2}-\kk^{2}}\int\! dn_{\perp} \left [ \int_{0}^{\infty}\! \frac{dy}{2\pi} e^{-\kk y\r} (-2i y + y^2)^{\frac{d-4}{2}} \htilde^{\mu\nu}(k) \right ] \ ,
\end{align}
where the two terms correspond to the left and right contours in Fig. \ref{fig:deformedcontour}, with $z=\pm 1 + i y$, respectively.

Since we are interested in the behavior of the gravitational field toward future null infinity, we can restrict to $t  >0$ and $\r$ being largest scales in the problem, with the retarded time $u=t \, -\r$ fixed. We can evaluate the \textit{asymptotic expansion} at large-$\r$ of the $y$ integrals (the integrals in the brackets): such integrals are dominated by the small-$y$ \textit{region} and their asymptotic expansions are fixed by integrating the expansion of the non-exponential part of the integrand around $y \simeq 0$ (\textit{Watson's lemma}). We therefore expand the source $J^{\mu\nu}(k)$ of the $y$ integral around $y=0$. A simple rescaling of $y$ then turns it into a series expansion in $\r^{-1}$. Each $y$ numerator factor pulls down a factor of $(\kk \r)^{-1}$.
These manipulations are equivalent to a multipolar expansion around the observer's direction. 

As we go to higher orders in the $1/\r$ expansion, the $\kk$ integration becomes more and more (infrared) divergent. This signals the importance of the Coulomb region, which we mentioned in the previous section. Indeed, beyond the leading order contribution in the $1/\r$ expansion, we expect that the contributions from the Coulomb region show an ultraviolet divergence, which cancels the corresponding infrared divergence in the radiation region. In Sec.~\ref{subsec:coulombregion}, we will show that the Coulomb region does not contribute at leading order in $1/\r$.

We still need to evaluate the integrals over the components $n_{\perp}$, $\kk$ and $\omega$ that parameterize the graviton momentum (cf. Eq.~\eqref{eq:parameterizationk}).  Upon expanding in $y$, the term-by-term integral over $n_{\perp}$ can be evaluated using the generating function
\begin{equation}\label{eq:perpintegraliden}
\int_{S^{n}} dn \, e^{a\cdot n}=(2 \pi )^{n/2} a^{\frac{2-n}{2}} I_{\frac{n-2}{2}}(a) \ ,
\end{equation}
which is valid for a vector $n$ parametrizing the unit $n$-sphere. 
In our case, $n_\perp$ is further constrained to be orthogonal on $t^\mu$ and $r^\mu$; the relevant projector enforcing this transversality is
\begin{equation}
    \eta_\perp^{\mu \nu} = \eta^{\mu \nu} - t^\mu t^\nu + r^\mu r^\nu\ \ ,
\end{equation}
and the integral is a function of only $a^\mu_\perp = \eta_\perp^{\mu \nu} a_\nu$.
The integrals over $\kk$ and $\omega$ are more interesting. They depend on the details of the source $\htilde^{\mu\nu}(k)\sim J^{\mu\nu}(k)$ and are evaluated on a case-by-case basis.
At tree-level, the $\omega$ integral can be evaluated via the residue theorem, as the only relevant singularities in the lower-half plane are the one in the external (retarded) propagator. Beyond tree level, the integral over $\omega$ is complicated by the appearance of logarithmic branch-points at $\omega=\kk$ in $J^{\mu\nu}(k)$ in $d=4$. We leave the evaluation of these integrals to Sections~\ref{sec:leadingwaveform}~and~\ref{sec:sublead}. As a cross-check, we will also show there that the leading contribution in $\r^{-1}$ reproduces the waveform at future null infinity, originally obtained in Ref.~\cite{Cristofoli:2021vyo} via a saddle-point approximation.

\subsection{The Coulomb region}
\label{subsec:coulombregion}

In this region, the external graviton momentum is taken to be parametrically soft, $\sim \r^{-1}$; moreover, since we are interested in the metric at large distance towards future null infinity, we will assume that, while $\omega \ne \kk$, they are of 
the same order, $\omega \sim \kk$.
Under these assumptions the source $\htilde^{\mu\nu}(k)$ can be expanded in powers of $\omega$ and $\kk$. Displaying only the $\kk$ dependence, $\htilde^{\mu\nu}(k)$ schematically has the general form~\cite{Georgoudis:2023eke}  
\begin{equation}
\label{eq:generic_soft_expansion}
    \htilde^{\mu\nu}(k) \sim \underbrace{\frac{1}{\kk} \sum_{n=0}^{\infty} \kk^n \, \htilde^{\mu\nu}_{1,n}(k)}_{\text{soft exchange}} + \underbrace{\kk^{d-4} \sum_{m=0}^{\infty} \kk^m \, \htilde^{\mu\nu}_{2,m}(k)}_{\text{ultra-soft exchange}}\ ,
\end{equation}
The terms in the first sum receive contributions from both on-shell and off-shell graviton momenta. Those in the second bracket, written here in a dimensionally-regularized form, are sensitive to the long-range nature of the gravitational interactions, including infrared divergences at loop level. 
This expansion is reminiscent of the expansion implied by the classical soft theorems~Ref.~\cite{Sahoo:2021ctw}. At leading order in Newton's constant, the terms in the second sum reflect the $1/\r$ falloff of gravitational interactions. Beyond leading-order, they receive contributions from the infinite propagation of the initial state as well as hereditary (tail) effects.
%
They originate both from higher-order corrections to the source of $\htilde^{\mu\nu}$ and from the Fourier transform to impact parameter space. It is thus convenient to consider together the Fourier transforms of both $k$ and of the momentum transfer, as we will do below.

Since $\omega, \kk \sim \r^{-1}$ in the Coulomb region, the phase of the Fourier transform in Eq.~\eqref{eq:deformed_angular_contour_sec_3} must be expanded in the small quantity $\omega u$:
\begin{equation}
    e^{-i k \cdot x} =  e^{-i k \cdot x} \Big|_{t = \r}  + \mathcal{O}(\omega u)\ .
    \label{eq:expansion_of_phase}
\end{equation}
Subleading orders contribute beyond the leading order in the $1/\r$ expansion. Following the usual procedure provided by the method of regions in dimensional regularisation, we take the upper limit of the integration domain to $\lambda \to \infty$ after series expanding the integrand.

As mentioned earlier in this section, the $\omega$ integral is evaluated by contour deformation, picking up the contributions from the singularities of the source and of the external propagator. If such singularities fix $\omega \propto k$ (as it is the case for \textit{e.g.} poles or branch cuts starting at $\omega = \pm \kk$, or for the singularities introduced by the eikonal matter propagators), the $\kk$ has a finite contribution which is dominated by $\kk \sim \r^{-1}$. On the other hand, if the singularities lie far off in the complex plane, the $\omega$ integration gives a contribution which is exponentially suppressed as we take $\r$ large by the Fourier phase factor.

In the following, we will focus only on the leading and subleading soft contributions in Eq.~\eqref{eq:generic_soft_expansion}. We will study both terms at all orders in the coupling. We find that only the subleading order in the soft expansion contribute to peeling violation.
In particular, we will see that the leading $\kk^{-1}$ contribution satisfies non-trivially the peeling conditions~\eqref{eq:sachssmooth} and that the leading term in the ultra-soft exchange region reproduces at leading order in $G$ the peeling violation~\eqref{eq:leadingpeelingviolation}, identified in Ref.~\cite{Christodoulou:2002}.

In two-body scattering, the source is
\begin{equation}
\begin{split}
    \label{eq:momentum_to_ips}
    J^{\mu \nu}(k,b_1,b_2) = \frac{1}{(2\pi)^{d-2}} & \!\int\! d^d q_1 d^d q_2\, \delta(2{\bar p}_1 \cdot q_1) \delta(2{\bar p}_2 \cdot q_2)\, \\
    &\times e^{i q_1\cdot b_1 + i q_2 \cdot b_2}\, \delta^d(q_1 + q_2 - k) J^{\mu \nu}(q_1, q_2)\ ,
\end{split}
\end{equation}
where the external momenta ${\bar p}_{1,2}$, \cite{Luna:2017dtq}
\begin{align}
{\bar p}_1 = p_1 - \frac{1}{2} q_1
\quad \textrm{and}\qquad
{\bar p}_2 = p_2 - \frac{1}{2} q_2 \ ,
\end{align}
are chosen so that the integrals in $J^{\mu \nu}(q_1, q_2)$ exhibit a clean separation of the classical orders.
$J^{\mu \nu}(q_1, q_2)$ is the final-state graviton one-point function in momentum space, and the two regions correspond to the following rescalings~\cite{Georgoudis:2023eke}: 
\begin{align}
\text{SE}:~(k^\mu, q_i^\mu)\sim (\r^{-1}, 1)\ ,
\qquad\qquad
\text{U-SE}:~(k^\mu, q_i^\mu)\sim (\r^{-1}, \r^{-1})\ .
\label{eq:scaling}
\end{align}
SE and U-SE stand for \emph{soft-exchange} and \emph{ultra-soft-exchange}, respectively. 
As usual in the method of regions in dimensional regularization, the integrals are evaluated by expanding at small $\omega$ in each region and integrating over the entire integration domain.

To illustrate the appearance of the two sums in Eq.~\eqref{eq:generic_soft_expansion} from the Fourier transform of the momentum transfer, we consider the simplest integral encountered in the computation of the source $J^{\mu\nu}(k, b_1, b_2)$ at leading order in Newton's constant~\cite{Cristofoli:2021vyo} ($b=b_1-b_2$ and we ignore an overall phase $e^{i b_2 \cdot k}$):
\begin{equation}
    \!\int\! \frac{d^d q_1}{(2\pi)^{d-2}} \, \delta(2p_1 \cdot q_1) \delta(2p_2 \cdot (k-q_1))\, \frac{e^{i b \cdot q_1}}{-q_1^2} = -\frac{\left(\frac{2 z}{-b^2}\right)^{\frac{d}{2}-2} K_{\frac{d}{2}-2}\!\left(z\right)}{2 m_1 m_2 \sqrt{\gamma^2-1}(4 \pi)^{\frac{d}{2}-1}}\ ,
\end{equation}
where $z= \sqrt{-b^2} p_2\cdot k/(m_2 \sqrt{\gamma^2-1})\propto \omega$ and $K_{\alpha}(z)$ is a modified Bessel function of the second kind. 
This function admits an asymptotic expansion around $z=0$, which is given in terms of two \textit{regions}:
\begin{equation}
\label{eq:besselexpansion}
    \left(2z\right)^{\alpha} K_{\alpha}\!\left(z\right) = \sum_{n=0}^{\infty} \frac{e^{-i \pi  n}}{2^{2n+1} \Gamma (n+1)} \left[2^{2 \alpha} \Gamma (\alpha -n)+ z^{2 \alpha} \Gamma (-n-\alpha ) \right] z^{2 n}\ .
\end{equation}
One can shown that the two sums correspond to expanding the integrand according to the scaling in Eq.~\eqref{eq:scaling}.

\subsection{Ansatz for metric}
\label{metricansatz}

Having discussed strategies for evaluating the asymptotic expansion of the metric sourced by a scattering process (which generates the source $J^{\mu\nu}$), let us now discuss strategies for streamlining the calculation of the Newman-Penrose scalars, which will serve as a diagnostic of asymptotic symplicity. 

As discussed, gauge invariance of the Einstein-Hilbert action coupled to the source $J^{\mu\nu}$ requires the source to be transverse. We thus parametrize it in terms of manifestly-transverse tensors. At lowest order in the graviton momentum, there are six such tensors, 
\begin{equation}
\begin{split}
    \label{eq:gauge_invariant_waveform}
    W_h(k) &= \varepsilon_{h, \mu \nu} J^{\mu \nu}(k) \ , \\
    &= \varepsilon_{h, \mu \nu} \sum_{i = 1}^6 \alpha_i \Big[(v_{i,1}\cdot k) v_{i,2}^\mu - (v_{i,2}\cdot k) v_{i,1}^\mu\Big]\Big[(v_{i,3}\cdot k) v_{i,4}^\nu - (v_{i,4}\cdot k) v_{i,3}^\nu\Big]\ ,
    \end{split}
\end{equation}
where the $v_i^\mu$ vectors are drawn from among the three external vectors $u_1^\mu$, $u_2^\mu$, and $b^\mu$ and the coefficients $\alpha_i$ depend on the kinematic variables of the scattering process.
${\cal O}(k^2)$ tensors can also be included but, as we will see later, they are not essential for our analysis.

The momentum-space metric follows by multiplying it with the conjugate polarization tensor and summing over all the physical polarizations,  
\begin{equation}\label{eq:onshellmetric}
    \htilde^{\mu \nu}(k) = \sum_h \varepsilon_{h}^{* \mu \nu} W_h(k) = \mathcal{T}^{\mu \nu}_{\mu^\prime \nu^\prime} J^{\mu^\prime \nu^\prime}(k)\ ,
\end{equation}
and the momentum-space linearized Weyl tensor is 
\begin{equation}\label{eq:weyltensorformlin}
\Ctilde^{\mu\nu\rho\sigma}(k)=-\frac{1}{2}k^{[\mu}k^{[\sigma}\htilde^{\nu]\rho]}(k)-\textrm{traces}+\textrm{non-linear} \ ,
\end{equation}
where the non-linear terms are discussed in Sec.~\ref{sec:nonlinearitiesetc}. The manifest transversality of the source in Eq.~\eqref{eq:gauge_invariant_waveform} implies that the reference vector drops out of Eq. (\ref{eq:weyltensorformlin}). 

The coefficients $\alpha_i$ turn out to be finite throughout the radiation region, $k^\mu \sim \r^0$, a property that 
we will leverage in later sections. 
To see this, recall that singularities of amplitudes (and of Green's functions) correspond to physical processes and, as such, can be understood using a combination of unitarity arguments and analytic continuation~\cite{Coleman:1965xm}.
In two-body scattering, the relevant invariants appearing in the impact-parameter-space form of the amplitude are $u_{1,2}\cdot k$, $b\cdot k$, $b^2$, $u_1\cdot u_2$, and $k^2$ (with the latter vanishing on-shell). 
The large distance expansion in the radiation region, corresponds to 
an essentially null momentum $k^\mu$, aligned with the null direction connecting the scattering event to the observer. 
In this limit, no singularities can arise. Indeed, the $u_i^\mu$ are time-like so divergences in $u_{i}\cdot k$ reduce to the standard soft singularities in $\omega$. Furthermore, no singularities can occur in $k\cdot b$, as the observer may lie anywhere on the celestial sphere and the impact parameter remains arbitrary. 
Moreover, since none of the tensor structures in Eq.~\eqref{eq:gauge_invariant_waveform} vanish as $z\rightarrow 1$, it follows that the coefficient functions $\alpha_i$ remain finite in this limit. These expectations can be confirmed explicitly from the known LO and NLO post-Minkowskian scattering waveforms~\cite{Kovacs:1977uw, Jakobsen:2021smu, Herderschee:2023fxh, Brandhuber:2023hhy, DeAngelis:2023lvf,Georgoudis:2023lgf, Bini:2023fiz,Brunello:2024ibk}.

The momentum-space Newman–Penrose scalars display richer angular dependence than the metric, arising both from explicit momentum factors and from their contractions with the null tetrad. This angular dependence can (and does) have nontrivial consequences on the Fourier-transform to position space.
To expose it, it is convenient to use spinor variables for the Weyl tensor and the tetrad instead of the representation in Eqs.~\eqref{eq:NPscalars}, \eqref{eq:tetrad_props} and \eqref{eq:metric_tetrad}.
In $d=4$, the Weyl tensor can be decomposed as
\begin{equation}
    C_{a \dot{a} b \dot{b} c \dot{c} d \dot{d}} = \Psi_{a b c d} \epsilon_{\dot{a} \dot{b}} \epsilon_{\dot{c} \dot{d}} + \bar{\Psi}_{\dot{a} \dot{b} \dot{c} \dot{d}} \epsilon_{a b} \epsilon_{c d} \ ,
\end{equation}
where $ \epsilon_{a b}$ and $\epsilon_{\dot{a} \dot{b}}$ are two-dimensional Levi-Civita symbol, $\Psi_{a b c d}$ is a fully symmetric tensor, $\bar{\Psi}_{\dot{a} \dot{b} \dot{c} \dot{d}} = (\Psi_{a b c d})^*$ and are the self-dual and anti-self-dual parts of the Weyl tensor, respectively. In the same way, the null tetrad (\ref{eq:metric_tetrad}) can be expressed in terms of spinors:
\begin{equation}
    N_{a \dot{a}} = \lambda_a \widetilde{\lambda}_{\dot{a}}\ , \quad L_{a \dot{a}} = \rho_a \widetilde{\rho}_{\dot{a}}\ , \quad \varepsilon_{-, a \dot{a}} = \lambda_{a} \widetilde{\rho}_{\dot{a}}\ , \quad \varepsilon_{+, a \dot{a}} = \rho_a \widetilde{\lambda}_{\dot{a}}\ ,
\end{equation}
where $\langle \lambda \rho \rangle = \sqrt{2} = [\rho \lambda]$. Once this basis is fixed, we can also decompose the spinor associated to the on-shell massless momentum $k^\mu$ in the basis given by $\lambda_a$ and $\rho_a$ (and their conjugates):
\begin{equation}
    \label{eq:4D_k_on-shell}
    \begin{split}
        k_{a \dot{a}} & = \frac{\omega}{\sqrt{2}} \xi_{a} \widetilde{\xi}_{\dot{a}}\ ,                                    \\
        \xi_{a}        = \sqrt{1-z}\, \lambda_a + e^{- i \phi} \sqrt{1+z}\, \rho_a\ , &                                   \qquad
        \xi_{\dot{a}}  = \sqrt{1-z}\, \widetilde\lambda_{\dot{a}} + e^{+i \phi} \sqrt{1+z}\, \widetilde\rho_{\dot{a}}\ .
    \end{split}
\end{equation}
With this parametrization, the Newman-Penrose scalars in Eq.~\eqref{eq:NPscalars} become 
\begin{equation}
    \label{eq:NP_scalars}
    \begin{split}
        \Psi_{4} & 
        =2 \Psi_{a b c d} \lambda^{a} \lambda^{b} \lambda^{c} \lambda^{d} \ , \qquad
        \Psi_{3} 
        = 2\Psi_{a b c d} \lambda^{a} \lambda^{b} \lambda^{c} \rho^{d} \ ,                 
        \Psi_{2} 
        = 2\Psi_{a b c d} \lambda^{a} \lambda^{b} \rho^{c} \rho^{d} \ ,  \\
      & \qquad \qquad
        \Psi_{1} 
        = 2\Psi_{a b c d} \lambda^{a} \rho^{b} \rho^{c} \rho^{d} \ ,                        
        \Psi_{0} 
        =2 \Psi_{a b c d} \rho^{a} \rho^{b} \rho^{c} \rho^{d} \ .
    \end{split}
\end{equation}
As promised, the angular dependence of the graviton momentum is explicit, Eq.~\eqref{eq:4D_k_on-shell}. 
This is a consequence of the fact that the spinor basis is only two-dimensional, allowing us to decompose the graviton momentum while at the same time explicitly solving the null condition for the tetrad.

\section{The peeling property by region}
\label{sec:peelinganalysis}

Using the methods outlined in the previous sections, we explicitly evaluate the asymptotic expansion of the Fourier transform~\eqref{eq:massageFT}.
We discuss separately the Coulomb and radiation regions, focusing in Secs.~\ref{leadingsoft}, \ref{sec:leading_Coulomb} and \ref{sec:leadingwaveform} on terms which are at rational in $\kk$ and $\omega$. 
These originate from the soft-exchange part~\eqref{eq:scaling} of the Coulomb region~\eqref{eq:k_regions}, and from the terms which are analytic in $k^2$ in the radiation region~\eqref{eq:k_regions}. 
We will prove that such terms satisfy Sachs's peeling property~\eqref{eq:sachssmooth} as a consequence of nontrivial cancellations coming from the angular integration. 
We then discuss the remaining contributions -- the ultra-soft-exchange part~\eqref{eq:scaling} of the Coulomb region~\eqref{eq:k_regions} in Sec.~\ref{Coulomb_ultra_soft} and the non-analytic terms in $k^2$ in the radiation region~\eqref{eq:k_regions} in Sec.~\ref{sec:sublead}.
The former yields violation of the peeling property already identified at leading order in Newton's constant
in Refs.~\cite{Christodoulou:2002}. The latter, which we discuss in Sec.~\ref{sec:sublead},  introduce a new stronger violation.

\subsection{Asymptotic behavior of the waveform: the gravitational memory}
\label{leadingsoft}

The leading terms in the soft expansion of S-matrix elements (and more generally in-out quantities) are known from Weinberg's classic analysis~\cite{Weinberg:1965nx}.
These results have been extended to classical regime, including the study of the soft expansion of waveforms~\cite{Laddha:2018rle,Laddha:2018myi,Laddha:2018vbn,Sahoo:2018lxl, Laddha:2019yaj,Saha:2019tub,Sahoo:2021ctw}. 
While these results were originally derived for on-shell gravitons, they are not affected by the off-shell nature of the gravitons considered here. The difference, 
\begin{equation}
    \frac{1}{(p_i + k)^2 -m_i^2} = \frac{1}{2 p_i \cdot k + k^2} \simeq \frac{1}{2 p_i \cdot k}\left(1 +{\cal O}\left(\frac{k^2}{2 p_i\cdot k}\right)\right) \ ,
\end{equation}
is on the one hand subleading in the soft limit and, on the other, it is analytic.  These properties allow us to ignore the difference at leading order. Direct power counting indicated that the off-shell deformation contributes only suppressed terms in the large-$\r$ expansion; moreover, as we will see, the fact that they are proportional to the off-shellness $k^2$ will also allow us to ignore them for the study of peeling properties. The leading soft behavior of the amplitude captures the information about asymptotic large retarded times $|u|$, \textit{i.e.} the so-called \textit{memory effect}.

The soft theorem for the two-body-scattering in-in correlator is\footnote{Such a soft theorem for the two-body scattering can be inferred \textit{e.g} from the result of Ref.~\cite{Saha:2019tub}: the leading soft theorem is given by 
\begin{equation}
    J^{\mu \nu} (k) = i \frac{\kappa^2}{2} \sum_{i=1}^{n+m} \frac{p_i^\mu p_i^\nu}{p_i \cdot k + i \epsilon}\ ,
\end{equation}
where $n$ and $m$ are the number of incoming and outgoing particles, respectively (either massive or massless; for the incoming particles the $p_i$ are intended as \textit{minus} their momenta). As our focus on the two-body scattering, we have to sum over the momenta of the incoming and outgoing massive states. During the scattering process, a certain amount of radiation is emitted and its total momentum is fixed by momentum conservation. Requiring gauge invariance to hold in the soft limit, fixes the fifth term to be of the form in Eq.~\eqref{eq:leading_soft}, where $\rho(K)$ is the energy spectrum, given by modulus squared of the waveform (see Ref.~\cite{Georgoudis:2025vkk} for a recent discussion).
Note also that we work in a mostly-minus metric convention while Ref.~\cite{Saha:2019tub} has the opposite, mostly-plus metric convention.
}
\begin{align}
    \label{eq:leading_soft}
        J^{\mu \nu} (k) &= i\frac{\kappa^2}{2} \left[-\frac{p_1^\mu p_1^\nu}{p_1 \cdot k - i \epsilon} - \frac{p_2^\mu p_2^\nu}{p_2 \cdot k - i \epsilon} + \frac{p_1^{\prime \mu} p_1^{\prime \nu}}{p_1^\prime \cdot k + i \epsilon} + \frac{p_2^{\prime \mu} p_2^{\prime \nu}}{p_2^\prime \cdot k + i \epsilon} \right.
        \nonumber\\[3pt]
        &\qquad\quad \; \left.+ \!\int\! d^d K \rho(K) \frac{K^{\mu} K^{\nu}}{K \cdot k + i \epsilon}\right] 
        + \mathcal{O}(\kk^0, \kk^{d-4}) \ , 
\end{align}
where $p_{1,2}^\mu$ are the two incoming matter momenta, $p_{1,2}^{\prime \mu}$ are (exact) outgoing momenta, and $K^\mu$ is the four-momentum carried by the emitted radiation during the scattering process. The outgoing matter momenta are determined by 
the details of the interactions between the two bodies, and we will not need their precise form, which is known only perturbatively. For example, at leading order in Newton's constant 
\begin{equation}
    \begin{split}
        p_1^{\prime \mu} &= m_1 u_1^\mu + \Delta p^\mu\ , \quad
        p_2^{\prime \mu}  = m_2 u_2^\mu - \Delta p^\mu\ , \quad
        K^\mu  = 0\ ,\\
        \Delta p^\mu &=  - 2\kappa^2 m_1 m_2 \frac{\left(\gamma ^2- \frac{1}{d-2}\right) \Gamma\left[\frac{d}{2}-1\right]}{16 \pi^{\frac{d-2}{2}} \sqrt{\gamma ^2-1} (-b^2)^{\frac{d-2}{2}}} \, b^\mu \, \overset{d=4}{=} \, \frac{2G m_1 m_2}{b^2} \frac{2\gamma^2-1}{\sqrt{\gamma^2-1}}\,  b^\mu\ .\\
    \end{split}
\end{equation}
More generally, momentum conservation relates the incoming and outgoing momenta as 
\begin{equation}
    p_1^\mu + p_2^\mu = p_1^{\prime \mu} + p_2^{\prime \mu} + \!\int\!d^d K \rho(K) K^\mu \ .
\end{equation}

It is convenient to introduce the shorthand notation $Q_i^\mu$'s to collectively denote all incoming and outgoing momenta, and $s_i$ as the signs in front of each term in Eq.~\eqref{eq:leading_soft} ($+1$ for the incoming momenta, $-1$ for the outgoing ones). The leading term in the soft expansion of the source is then
\begin{equation}
    \begin{split}
        J^{\mu \nu} (k) &=  i \frac{\kappa^2}{2} \sum_{i=1}^{5} s_i 
\frac{Q_i^{\mu} Q_i^{\nu}}{Q_i \cdot k + s_i i \epsilon} + \mathcal{O}(\kk^0, \kk^{d-4})\ ,
    \end{split}
\end{equation}
The momentum space metric is given by contraction of the source with the projector~${\cal T}$ in Eq.~\eqref{eq:projector}; in this contraction, the reference vector $\zeta^\mu$ drops out due to momentum conservation. 

It is computationally convenient to evaluate the integrals in the rest frame of $Q_i$
\begin{align}
\label{eq:restframe}
 Q_i^\mu = \sqrt{Q_i^2}\; t^\mu \ ,
\end{align}
and restore the space-like components of $Q_i$ with a suitable boost. 
The relevant integral is therefore
\begin{equation}
\label{eq:leadingsoft_integral}
    \begin{split}
        I_i &= \int\! \frac{d^{d}k}{(2\pi)^{d}}  \frac{e^{-ik\cdot x} }{[(k_{0}+i\epsilon)^2-\vec{k}^{2}] (Q_i \cdot k + s_i i \epsilon)} \\
        &= \frac{1}{\sqrt{Q_i^2}} \int_{0}^{\infty}\! \frac{d\kk}{2\pi} \int_{-\infty}^{+\infty}\!\frac{d\omega}{2\pi}\frac{\kk^{d-2} e^{-i\omega t} }{[(\omega+i\epsilon)^{2}-\kk^{2}] (\omega+ s_i i \epsilon)}  \int\! dn \, e^{- i \kk n \cdot x}\\
        &= \frac{i}{\sqrt{Q_i^2}} \int_{0}^{\infty}\! \frac{d\kk}{2\pi}\, \kk^{d-4} \left(\cos\!\left(\kk t\right) - \delta_{1,s_i}\right) J_{\frac{d-3}{2}}(\kk \r) \left(2\pi \kk \r\right)^{\frac{3-d}{2}} \ ,\\
    \end{split}
\end{equation}
where the Bessel function is the result of the angular integration, cf.~Eq.~\eqref{eq:perpintegraliden} and $\delta$ is Kronecker's delta of the specified arguments.
The remaining integral is absolutely convergent for $2<d<4$, as it is a nontrivial function of $t/\r$.\footnote{In this computation, the use of dimensional regularization is convenient. The computation can be performed equally well in $d=4$. The integral in $\omega$ is performed using Cauchy's theorem (including the poles from the matter propagator and the external retarded graviton), followed by an asymptotic expansion of the angular integration, as shown in Sec.~\ref{sec:radiationreg}. At this point, we can consider the limit $u\to \pm\infty$. This step can be also performed modifying the integration contour of $\kk$, to make it exponentially suppressed for large values of $\kk$. Finally, Watson's lemma will guarantee that the leading term is captured by the asymptotic expansion of the current as $\kk \to 0$.}
We can verify {\it a posteriori} that the integrals vanish if the retarded external graviton propagator is canceled by terms proportional to $k^2$. Therefore, we can consistently ignore the contributions of such off-shell terms, as they also are subleading in the soft limit, and we will do so in the following.

As mentioned, we boost the result from rest frame to a generic one to restore the complete dependence on the vectors $Q_i$. This effectively amounts to replacing the time $t$ and norm $\r$ as
\begin{equation}
    t = \frac{Q_i \cdot x}{\sqrt{Q_i^2}}\ ,\quad \r = \frac{\sqrt{(Q_i \cdot x)^2 - Q_i^2 x^2}}{\sqrt{Q_i^2}}\ .
\end{equation}
Fixing $d=4$ and carrying out the $\kk$ integration in Eq.~\eqref{eq:leadingsoft_integral}, yields
\begin{equation}
    I_i 
    =
    i \frac{\delta_{1, -s_i} \Theta(-u) - \delta_{1, s_i} \Theta(u)}{4\pi \sqrt{(Q_i \cdot x)^2 - Q_i^2 x^2}}\ ,
    \label{eq:scalarIntFinal}
\end{equation}
where $\delta_{1, -s_i}=1-\delta_{1, s_i}$.
This recovers the well-known result that the leading term in the soft expansion captures the gravitational-wave memory, obtained here by approaching ${\cal I}^+$ from finite distances.\footnote{Analogous results for scalar waves were obtained \cite{Donnay:2022ijr} by directly solving the Klein-Gordon equation.}
We note that this integral depends on $Q_i$ and $x$ via the combination $(Q_i \cdot x)^2 - Q_i^2 x^2$; this observation will be important shortly.

This non-vanishing integral receives contributions from zero-frequency internal gravitons. Indeed, as we discussed, canceling the external graviton propagator leads to vanishing 
integrals, so the graviton propagator in $I_i$ is effectively cut. Therefore, contributions from the residue of the eikonal propagator are proportional to (the square of) an on-shell three-point amplitude, which has support only on zero-frequency gravitons.

The linearized Weyl tensor that is determined by the leading-soft momentum-space metric is, using Eq.~\eqref{eq:weyltensorformlin},
\begin{equation}
    \label{eq:leadingsoft_weyl}
    \widetilde{C}^{\mu\nu\rho\sigma}(k) = -\frac{i \kappa^2}{4} \sum_{i=1}^{5} \frac{s_i}{Q_i \cdot k + s_i i \epsilon} \left(k^{[\mu} Q_i^{\nu]} k^{[\sigma} Q_i^{\rho]} + Q_i^2 \frac{k^{[\mu} \eta^{\nu ] [\sigma} k^{\rho]}}{d-2} \right)+ \mathcal{O}(\kk^0, \kk^{d-4})\ .
\end{equation}
Its tracelessness and on-shell transversality are a consequence of momentum conservation. The two-tensor integral needed to find the position-space linearized Weyl tensor is obtained taking the derivative with respect to $x^\mu$ of $I_i$,
\begin{equation}
    I_i^{\mu_1 \cdots \mu_n} \equiv \int\! \frac{d^{d}k}{(2\pi)^{d}} \frac{e^{-ik\cdot x}\, k^{\mu_1} \cdots k^{\nu_n}}{[(k_{0}+i\epsilon)^2-\vec{k}^{2}] (Q_i \cdot k + s_i \, i \epsilon)} = i^n \frac{\partial^n I_i }{\partial x^{\mu_1} \cdots \partial x^{\mu_n}}\ .
\end{equation}
leading to
\begin{equation}
    I^{\mu \nu}_i = i \frac{\delta_{1,-s_i} \Theta(-u) - \delta_{1,s_i} \Theta(u)}{4 \pi  \left[(Q_i\cdot x)^2-Q_i^2 x^2\right]^{3/2}} \left[ Q_i^2 \eta ^{\mu \nu}-Q_i^{\mu } Q_i^{\nu }+3 \frac{\left(Q_i\cdot x\, Q_i^{\mu }-Q_i^2 x^{\mu }\right) \left(Q_i\cdot x\, Q_i^{\nu }-Q_i^2 x^{\nu }\right)}{(Q_i\cdot x)^2-Q_i^2 x^2} \right]\ .
\end{equation}
In particular, we notice that, in $d=4$, such integrals are transverse with respect to $Q_i$, i.e.  $Q_{i \mu_j} I_i^{\mu_1 \cdots \mu_n} = 0$. This is a consequence of the earlier observation that the scalar integral depends solely on the combination $(Q_i\cdot x)^2-Q_i^2 x^2$. 
We also note that both the scalar integral -- and hence the metric --  features a non-vanishing term for negative (large) values of the retarded time at leading order in $\r^{-1}$. Such contributions are subtracted by the subtraction at $t\to -\infty$ in Eq.~\eqref{eq:h}.

Putting together all the pieces that contribute to the Weyl tensor and subsequently constructing the Newman-Penrose scalars, we find 
{
\allowdisplaybreaks
\begin{align}
        \Psi_4 &= \frac{\kappa^2}{2} \sum_{i=1}^5 \left[-\delta_{-s_i} \Theta(-u) + \delta_{s_i} \Theta(u)\right] Q_i^4 \frac{3 (\varepsilon_- \cdot Q_i)^2}{16 \sqrt{2} \pi  (L\cdot Q_i)^5 \r^3}+\mathcal{O}\!\left(\r^{-4}\right)\ ,\notag\\
        \Psi_3 &= \frac{\kappa^2}{2} \sum_{i=1}^5 \left[-\delta_{-s_i} \Theta(-u) + \delta_{s_i} \Theta(u)\right] Q_i^4 \frac{3 (\varepsilon_-\cdot Q_i)}{32 \sqrt{2} \pi  (L\cdot Q_i)^4 \r^3}+\mathcal{O}\!\left(\r^{-4}\right)\ ,\notag\\
        \Psi_2 &= \frac{\kappa^2}{2} \sum_{i=1}^5 \left[-\delta_{-s_i} \Theta(-u) + \delta_{s_i} \Theta(u)\right] Q_i^4 \frac{1}{32 \sqrt{2} \pi  (L\cdot Q_i)^3 \r^3}+\mathcal{O}\!\left(\r^{-4}\right)\ ,\\
        \Psi_1 &= - \frac{\kappa^2}{2} \sum_{i=1}^5 \left[-\delta_{-s_i} \Theta(-u) + \delta_{s_i} \Theta(u)\right] Q_i^4 \frac{3 u (\varepsilon _+\cdot Q_i)}{64 \sqrt{2} \pi  (L\cdot Q_i)^4 \r^4}+\mathcal{O}\!\left(\r^{-5}\right)\ ,\notag\\
        \Psi_0 &= \frac{\kappa^2}{2} \sum_{i=1}^5 \left[-\delta_{-s_i} \Theta(-u) + \delta_{s_i} \Theta(u)\right] Q_i^4 \frac{3 u^2 (\varepsilon_+\cdot Q_i)^2}{64 \sqrt{2} \pi  (L\cdot Q_i)^5 \r^5}+\mathcal{O}\!\left(\r^{-6}\right)\notag\ ,
\end{align}
}
which is consistent with Sachs's peeling property~\eqref{eq:sachssmooth}.\footnote{The appearance of the vector $L$ in the denominator is a consequence of the expansion at large $\r$. Indeed, writing $x^\mu = t \, t^\mu + \r r^\mu = u t^\mu +\sqrt{2} \r L^\mu$, the natural denominator becomes $(Q_i\cdot x)^2-Q_i^2 x^2 = \r^2 (L\cdot Q_i)^2(1+ {\cal O}(u/\r)$.}
The appearance of higher powers of $1/\r$ in $\Psi_0$ and $\Psi_1$ may appear mysterious. For this contribution, it is a consequence of momentum conservation and projection onto the null tetrad~\eqref{eq:metric_tetrad}, as we have derived this by making explicit use of the simple form of the leading terms in the soft expansion.
We will systematically understand the suppression of the $\Psi_1$ and $\Psi_0$ scalars with respect to the other Newman-Penrose scalars in Sec.~\ref{sec:leadingwaveform}, as a consequence of the angular integration.

\subsection{Coulomb region}
\label{sec:coulomb_region}

In this section, we restrict to the leading order in the Coulomb region, which we expect to be a part of the memory. As we discussed, the Coulomb region is defined to contain outgoing gravitons with extremely soft, ${\cal O}(|x|^{-1})$, momenta, so a soft expansion of the source is justified.

Before proceeding with the detailed calculation using the universal form of the soft expansion~\cite{Saha:2019tub}, let us discuss a scaling argument that indicates which NP scalars could potentially exhibit departure from the expected peeling property and the loop order at which this could occur.
To this end, it is useful to inspect the source in momentum-transfer space and recall that the classical limit specifies the scaling of its $L$-loop component, $J^{\mu\nu}_L$,  as
\begin{equation}
(k, q_i)\rightarrow \lambda (k, q_i)
\qquad
J^{\mu\nu}_L(\lambda k, \lambda q_i) = 
\lambda^{L-2} 
J^{\mu\nu}_L(k, q_i),
\end{equation}
where the scaling of an overall momentum conserving delta function, $\delta^{d}(q_1+q_2-k)$ is not included.
Then, position-space metric is given by the Fourier transforms over $k$ and $q_i$, together with the external graviton propagator. 

Restricting to the Coulomb and U-SE region, $\lambda = \r^{-1}$ we recall that, as discussed in Sec.~\ref{subsec:coulombregion}, to leading order we may ignore the Fourier phases; thus, the integration measure and additional factors bring an additional factor of $\lambda^{2d-4}$, leading altogether for the $L$-loop linearized Weyl tensor \eqref{eq:weyltensorformlin} to
\begin{equation}
\Ctilde_L^{\mu\nu\rho\sigma}(k, q_i)\Big|_{\substack{\text{Coulomb}\\\text{U-SE}}}  \longmapsto 
\lambda^{L+2d-4} \; \Ctilde_L^{\mu\nu\rho\sigma}(k, q_i)\Big|_{\substack{\text{Coulomb}\\\text{U-SE}}}  \ .
\end{equation}
We see that, in the Coulomb and U-SE region, the $L$-loop linearized Weyl tensor scales as
\begin{equation}
\label{eq:scalingX}
C_L^{\mu\nu\rho\sigma}(x)\Big|_{\substack{\text{Coulomb}\\\text{U-SE}}} \sim  \frac{1}{\r^{L+2d-4}} \ .
\end{equation}
Consequently, for $d=4$, we see that this region can depart from the expected peeling property only at tree level, $L=0$, and moreover that 
it can occur only for the NP scalar $\Psi_0$, which is expected to scale as $\r^{-5}$ if the peeling property holds. 
We note that this argument does not cover logarithmic contributions. Indeed, we will show that even for $\Psi_1$, we will reproduce a known logarithmic departure from the peeling conditions.

\subsubsection{Coulomb region: Leading-soft contribution}
\label{sec:leading_Coulomb}

Let us now focus on the leading terms in the soft expansion. Having already evaluated their Fourier transform without restricting to the Coulomb region, it is instructive to study the consequences of imposing this restriction.
Assuming that $u\ll \r$ and expanding the Fourier phase in 
$k_0 u \sim u/\r$ as discussed in Sec.~\ref{subsec:coulombregion}, we have
\begin{equation}
 {C}^{\mu\nu\rho\sigma}(x)  = - \int\! \frac{d^{d}k}{(2\pi)^{d}} \left.\frac{e^{-ik\cdot x} \, (1-i k_0 u +\dots)}{(k_{0}+i\epsilon)^2-\vec{k}^{2}} \right|_{t = \r} \widetilde{C}^{\mu\nu\rho\sigma}(k)\ .
\end{equation}

As before, we will focus on the scalar integral defining the metric.
Even though, because of the expansion around the null direction, the tensor integrals cannot be obtained by differentiation, they are a  simple generalization of the scalar integral that we discuss.
Keeping the same notation, Eq.~\eqref{eq:leadingsoft_integral} becomes
\begin{equation}
\label{eq:leadingsoft_integral_C}
    \begin{split}
        \left. I_i \right|_{t = \r} &= \int\! \frac{d^{d}k}{(2\pi)^{d}} \left. \frac{e^{-ik\cdot x} }{[(k_{0}+i\epsilon)^2-\vec{k}^{2}] (Q_i \cdot k + s_i i \epsilon)}\right|_{t = \r}  \\
        &= \frac{i}{\sqrt{Q_i^2}} \int_{0}^{\infty}\! \frac{d\kk}{2\pi}\, \kk^{d-4} \left(\cos\!\left(\kk \r\right) - \delta_{1,s_i}\right) J_{\frac{d-3}{2}}(\kk \r) \left(2\pi \kk \r\right)^{\frac{3-d}{2}} \ .\\
    \end{split}
\end{equation}
As for $t\ne \r$, the $\kk$ integral is absolutely convergent for $2<d<4$, with the dimensional regularization parameter $\epsilon = (4-d)/2$ acting as an UV regulator. The result is:
\begin{equation}
    \left. I_i \right|_{t = \r} = \frac{i\, \delta_{1, -s_i}\, \Gamma\!\left[\frac{d-3}{2}\right]}{4\sqrt{Q_i^2} \pi^{\frac{d-1}{2}} \r^{d-3}}\ .
\end{equation}
That is, $\left.I_i \right|_{t = \r}$ is nonzero only for the incoming particles.

The same result can be obtained by regularizing the integral with a cut-off $\lambda$ and evaluating it in $d=4$, as in Eq.~\eqref{eq:soft_hard_split}, but the interpretation of the result is more subtle.~\footnote{The method of region in dimensional regularization allow us to integrate $\kk$ in the interval $(0,+\infty)$, as long as we choose $d$ such that the integral is absolutely convergent. Such result is analytically continued to the complex $d$-plane. In alternative, we could have chosen $d=4$ and integrate $\kk$ in the interval $(0,\lambda)$, as in Eq.\eqref{eq:soft_hard_split}. At leading order in the Coulomb region, we can take the limit $\lambda \to + \infty$. The final result is expected to match asymptotically the analytic result, and the dependence on $\lambda$ to cancel between and Coulomb and the radiation region. Thus, in the cut-off-regulated integration, the $\lambda$-independent terms match the result in dimensional regularization, as we take $d\to 4$. Since the integral is not absolutely convergent, the integral computed in $d=4$, without a cut-off, will not give from the correct result.} 
To understand this subtlety, we evaluate the integral in dimensional regularization {\it and} with a cut-off:
\begin{equation}
    \left. I_i \right|_{t = \r} = \frac{i}{4\sqrt{Q_i^2} \pi^{\frac{d-1}{2}} \r^{d-3}} \left[ \delta_{1, -s_i}\, \Gamma\!\left[\frac{d-3}{2}\right] + \frac{\lambda ^{\frac{d}{2}-2} \sin\frac{\pi  d}{4}}{2^{\frac{d}{2}-3}\sqrt{\pi } (d-4)} + \mathcal{O} (e^{\pm i \lambda} \lambda ^{\frac{d}{2}-3}, e^{\pm 2 i \lambda} \lambda ^{\frac{d}{2}-3}) \right]\ .
\end{equation}
This shows explicitly that, with a cut-off in $d=4$, the integral has an extra (finite) term. Since this term is proportional to (a power of) $\lambda$ in generic dimensions, we expect that is cancels with an infrared finite term originating in the radiation region. Working in $d<4$, we automatically remove such term.

Moreover, in dimensional regularization, subleading orders in $\omega u$ yield only scaleless integrals (one may see this \textit{e.g.} by implementing the IBP relations for the integrals family in the first line of Eq.~\eqref{eq:leadingsoft_integral}, following Ref.~\cite{Brunello:2024ibk}), so the leading order in the expansion around the null direction connecting the scattering event and the observer gives the complete contribution of the leading soft theorem.

To summarize, our explicit calculation shows that \emph{only the incoming states contribute in the Coulomb region at leading order in the soft expansion}. 
It is not difficult to show that this conclusion holds, in fact, to all orders in the soft expansion. The singularities of any finite-order term 
are given by products of $Q_i\cdot k$ and, consequently, absence of pinches of the integration contour as $i\epsilon\rightarrow 0$ requires that they are all on the same side of the contour. 
We will now argue that, if these singularities are all in the lower-half plane, then the Fourier transform of such terms vanishes in dimensional regularization. To this end we will show that the Fourier transform in the Coulomb region of any homogeneous function of $k^\mu$ with singularities only in the lower-half plane vanishes in dimensional regularization.
Consider such a function, $f_R (k^\mu)$; its homogeneity allows us to organize it as
\begin{equation}
    f_R(k^\mu) = f_R(\omega, \kk, n) \equiv \kk^\alpha \tilde{f}_R ( \tilde{\omega}, n)\ , \qquad \omega \equiv \kk \, \tilde{\omega}\ ,
\end{equation}
which, because of our assumption that all the singularities of $f_R$ are in the lower-half plane, is analytic in the $\tilde{\omega}$ upper-half plane.
Thus, we find
\begin{equation}
\label{eq:nocontribfromLHP}
    \begin{split}
        \int \!d^d k e^{-ik\cdot x } f_R(k^\mu)\Big|_{\text{Coulomb}} &= \int \!dn \int_{-\infty}^{+\infty} \!d\tilde{\omega} \int_0^\infty d\kk\, \kk^{d-1+\alpha} e^{-i \kk \r (\tilde{\omega}- n \cdot r)} \tilde{f}_R ( \tilde{\omega}, n)\\
        &= \frac{\Gamma\left[d+\alpha\right]}{(i \r)^{d+\alpha}} \int \!dn \int_{-\infty}^{+\infty} \!d\tilde{\omega}\, \frac{\tilde{f}_R ( \tilde{\omega}, n)}{(\tilde{\omega} - n \cdot r)^{d+\alpha}} = 0\ ,
    \end{split}
\end{equation}
where setting $t=\r$ in the first equality is a consequence of the restriction to the Coulomb region. The singularity on the real axis can be made integrable by a careful choice of the dimensional regularization parameter $d$. The last equality follows from Cauchy's theorem, by deformation of the integration contour on the upper-half plane, where we assumed there are no singularities. In the lower-half plane, there are always two poles from the external retarded graviton propagator.

\subsubsection{Coulomb region: subleading-soft and peeling violation}
\label{Coulomb_ultra_soft}

We argued that, because gravitons in the Coulomb region have momenta $~\r^{-1}$, their contribution to the metric is governed by the soft expansion of the source. 
Having discussed the leading-order contributions in the previous section, we now proceed to discuss the first subleading contributions.
We will find agreement with the results of Refs.~\cite{Christodoulou:2002}. 

Similarly to the leading soft terms, the form of the subleading soft terms are universal~\cite{Saha:2019tub}, independent of the details of the scattering process, when expressed in terms of the incoming and outgoing momenta:\footnote{The form of the subleading soft theorem may not look covariant. On the other hand, all naively covariant deformations are all equivalent to this order in the soft expansion, \textit{e.g.} $\log k \cdot Q \sim \log \omega + \mathcal{O}(\omega^0)$.}$^,$\footnote{In the following equation, the sums run from the one to five, where fifth element is meant to be the sum over emitted radiation. Whenever the emitted radiation is appearing more then once, the integration weigted by the spectrum is to supposed to be overall, as in equation~\eqref{eq:leading_soft}.}
\begin{equation}
\label{eq:subleading_soft_from_sen_et_al}
    \begin{split}
        J^{\mu \nu}(k)\Big|_{\omega^0 \ln\omega} = &- \frac{\kappa^4}{32\pi} 
        \log(\omega - i \epsilon) F(Q_1, Q_2) \frac{Q_1 \cdot Q_2}{k \cdot Q_1} Q_1^\mu\, k \cdot Q_{[2}\, Q_{1]}^\nu + (1\leftrightarrow 2)\\
        & - \frac{\kappa^4}{32\pi} \log(\omega + i \epsilon) \sum_{a=1}^3 \sum_{\substack{b=1 \\b \neq a}}^3 F(Q_a^{\prime},Q_a^{\prime}) \frac{Q^\prime_a \cdot Q^\prime_b}{k \cdot Q^\prime_a} Q_a^{\prime \mu}\left(k \cdot Q^\prime_{[b}\, Q_{a]}^{\prime \nu}\right)\\
        &-\frac{\kappa^4}{64\pi}  \log(k^2 + i \epsilon) \sum_{a=1}^3 Q^{\prime}_a\cdot k \left[ \sum_{b=1}^3 \frac{Q^{\prime \mu}_aQ^{\prime \nu}_a}{k\cdot Q^{\prime}_a+i\epsilon} -  \sum_{b=1}^2 \frac{Q^{\mu}_a Q^{\nu}_a}{k\cdot Q_a+i\epsilon} \right]
    \end{split}
\end{equation}
where 
\begin{equation}
F(A, B) = \frac{\frac{3}{2}A^2 B^2 - (A\cdot B)^2}{((A\cdot B)^2-A^2 B^2)^{3/2}}
\label{eq:sub_logk}
\end{equation}
and $Q_3^{\prime \mu}=K^\mu$ is the total radiated energy during the scattering and the dots stand for the leading soft contributions.
These terms originate both from the off-shell momentum-space source as well as from the Fourier-transform to impact-parameter space, from the ultra-soft-exchange region.\footnote{On the last line we introduced $\ln(\omega+i\epsilon)\rightarrow \frac{1}{2} \ln(k^2+i\epsilon)$ because of the proportionality with the leading order soft factor and because of the consistency of the coefficients with Weinberg's exponentiation of IR divergences. We also note that the $i\epsilon$ deformations of singularities are such that the integration contour is not pinched between these logarithmic branch points and the poles of the external propagator.}

The soft theorem in Eq.~\eqref{eq:subleading_soft_from_sen_et_al} has been derived in $d=4$. This is a clear obstruction to performing this computation within dimensional regularization and the method of regions. Thus, in this case, we are forced to compute the contribution of the Coulomb soft region using cut-off regularization. Power counting suggests that such integrals are UV power-divergent as the cutoff is taken $\lambda \to +\infty$. As discussed below Eq.~\eqref{eq:soft_hard_split}, such divergences should cancel when the Coulomb and radiation regions are added up, leaving finite terms that depend on the parameters of the integral. Furthermore, we will ignore the retarded contributions ({\it i.e.} with denominators $(Q_i \cdot k + i \epsilon)$ and $\log(\omega + i \epsilon)$) because they vanish in dimensional regularization, as discussed around Eq.~\eqref{eq:nocontribfromLHP}. Thus, in other regularizations, we expect them to be UV sensitive and cancel completely against counterparts in the radiation region.

As discussed in Sec.~\ref{sec:asymptotic_expansion}, we must discuss the off-shell transversality of the source and, if absent, restore it through a gauge transformation.
Up to terms ${\cal O}(i\epsilon)$, the third line is proportional to the leading term in the soft expansion, which is gauge invariant. 
Moreover, we can easily check that the first two lines are automatically transverse even off-shell on their own. Thus, for our purpose, we may consider $J^{\mu\nu}$ in Eq.~\eqref{eq:subleading_soft_from_sen_et_al} as transverse off-shell in the Coulomb region.
Consequently, upon contraction with the projector \eqref{eq:projector}, all reference-vector dependence drops out from the linearized Weyl tensor. 
Finally, we should comment on the reality condition of the metric in the Coulomb region. Indeed, we can immediately check that
\begin{equation}
    J^{\mu \nu} (-k) \neq [J^{\mu \nu}(k)]^*\ ,
\end{equation}
and the metric looks complex. On the other hand, this is related to $i\pi$ terms, which are subleading in the soft expansion:
\begin{equation}
    \log(\omega \pm i \epsilon) = \text{Re}\left[\log(\omega \pm i \epsilon)\right] + \mathcal{O}(\omega^0)\ .
\end{equation}
Thus, we are going to ignore any $i \pi$ term originating from the evaluation of the logarithm along the branch cut.

The Fourier transform of Eq.~\eqref{eq:subleading_soft_from_sen_et_al} amounts to evaluating integrals of the type\footnote{\label{cutoffdefinition}
The shorthand notation introduced below for the integration is defined as 
\begin{equation}
    \int^\lambda\! \frac{d^4k}{(2\pi)^4} \equiv \int_{-\infty}^{+\infty}\!\frac{d\omega}{2\pi}\int_{0}^{\infty}\! \frac{d\kk}{2\pi} e^{-\frac{\kk}{\lambda^\prime}}\kk^2 \int\! dn \sim \int_{-\infty}^{+\infty}\!\frac{d\omega}{2\pi}\int_{0}^{\lambda}\! \frac{d\kk}{2\pi} \kk^2 \int\! dn \ .
\end{equation}
The last step should be understood as an asymptotic equality as we take the cutoff $\lambda, \lambda^\prime \to + \infty$ and the two differ by a finite rescaling. Moreover, to enforce the restriction to the Coulomb region, the cutoffs must be taken to depend on $\r$ as $\lambda = \frac{\Lambda}{\r}$, with $\Lambda \gg 1$.
}${}^,$\footnote{We inserted here $i\epsilon$ shifts in the linear propagators of Eq.~\eqref{eq:subleading_soft_from_sen_et_al} by demanding that the integration contour is not pinched between the pole of the propagator and the logarithmic branch point.}
\begin{equation}
    \int^\lambda\! \frac{d^4k}{(2\pi)^4} \left. \frac{e^{-ik\cdot x}}{(\omega+i \epsilon)^2 - \kk^2} \frac{\log(\omega - i \epsilon)}{(Q_i \cdot k - i \epsilon)} k^{\mu} \right|_{t = \r}\ ,
\end{equation}
for the metric, and
\begin{equation}
    \int^\lambda\! \frac{d^4k}{(2\pi)^4} \left. \frac{e^{-ik\cdot x}}{(\omega+i \epsilon)^2 - \kk^2} \frac{\log(\omega - i \epsilon)}{(Q_i \cdot k - i \epsilon)^{n-2}} k^{\mu_1} \cdots k^{\mu_n} \right|_{t = \r} \qquad (n=2,3)\ ,
\end{equation}
for the Weyl tensor, where the restriction to $t=\r$ is a consequence of expanding in the Coulomb region. We notice that, by computing the $\omega$ integration by Cauchy's theorem, we are sensitive only to the on-shell poles of the retarded graviton propagator because $t=\r>0$. Thus, all off-shell terms ($\propto k^2$) give vanishing contributions. 
One can show via direct evaluation in dimensional regularization that all the integrals with retarded matter propagators ({\it i.e.} with denominators $(Q_i \cdot k + i \epsilon)$ and $\log(\omega + i \epsilon)$) give vanishing integrals, see Eq.~\eqref{eq:nocontribfromLHP}.\footnote{As already briefly mentioned below Eq.~\eqref{eq:sub_logk}, this statement hides some subtleties. Indeed, by using a different regulator than dimensional regularization, one may find that the retarded contributions give a non-vanishing contribution. Such terms are, however, expected to cancel exactly with analogous regulator-dependent terms in the radiation region.} 
Thus, the details of the outgoing particles -- in particular the impulse -- do not contribute to the Coulomb region, recovering the conclusion of the scaling argument that the Coulomb (and U-SE) region is tree-level exact, from a different angle.

It is instructive to show the explicit result for the metric integral and the integral with two tensors powers appearing in the linearized Weyl tensor. In particular, the former is particularly challenging to compute for generic velocities, so we naturally restrict ourselves to the small-velocity expansion:
\begin{equation}
    Q\cdot t \gg Q\cdot r, Q_\perp^2\ ,
\end{equation}
where
\begin{equation}
    Q^\mu = (Q\cdot t) t^\mu - (Q\cdot r) r^\mu + Q_\perp^\mu\ .
\end{equation}
Thus, we find:
\begin{equation}
\label{eq:metric_integral_subleading}
    \begin{split}
        \int^\lambda\! \frac{d^4k}{(2\pi)^4} \frac{e^{-i\r k\cdot L}}{(\omega+i \epsilon)^2 - \kk^2} &\frac{\log(\omega - i \epsilon)}{(Q_i \cdot k - i \epsilon)} k^{\mu} = -\frac{[\log( e^{-\gamma_E} \lambda)]\lambda}{4 \pi ^2 \r (L\cdot Q)} L^\mu -\frac{\log( 2e^{\gamma_E} \r)}{4 \pi ^2 \r^2 (L\cdot Q)} L^\mu\\
        +& \frac{\log( 2e^{\gamma_E} \r)-1}{4 \pi ^2 \r^2 (L\cdot Q)} N^\mu +\frac{[\log (2e^{\gamma_E}\r)-2] (L\cdot Q-N\cdot Q)}{4 \pi  \r^2 L\cdot Q^2} L^\mu\\
        +&\frac{L\cdot Q-N\cdot Q}{4 \pi  \r^2 L\cdot Q^2} N^\mu+\frac{\log (2e^{\gamma_E}\r)-2}{4 \pi  \r^2 L\cdot Q^2}  Q_{\perp}^\mu + \dots \ ,
    \end{split}
\end{equation}
and
\begin{equation}
\label{eq:Weyl_integral_subleading}
    \begin{split}
        \int^\lambda\! \frac{d^4\kk}{(2\pi)^4} e^{-ik\cdot x} & \frac{\log(\omega - i \epsilon)}{(\omega+i \epsilon)^2 - \kk^2} \, k^{\mu} k^{\nu} \bigg|_{t = \r} = -\frac{ \left[2 \log \left(e^{-\gamma_E} \lambda \right) + 3\right]\lambda ^3}{2 \pi ^2 \r}L^\mu L^\nu\\
        +&\frac{\log \left(e^{-\gamma_E} \lambda \right) \lambda}{4 \pi^2 \r^3} \left(L^\mu L^\nu + N^\mu N^\nu + \eta^{\mu \nu}-2 L^\mu N^\nu-2 N^\mu L^\nu\right)\\
        +&\frac{3 \left(\eta^{\mu \nu} -2 L^\mu N^\nu-2 N^\mu L^\nu\right)+2 L^\mu L^\nu +6 N^\mu N^\nu}{32 \pi  \r^4}+\mathcal{O}\left(\lambda^{-1}\right)\ ,
    \end{split}
\end{equation}
where we can explicitly check that the second integral is traceless, \textit{i.e.} that the off-shell contributions integrate to zero. We provided the explicit result for the first integral, which contributes to the metric, to subleading order in the small-velocity expansion (the first three terms corresponding to the leading order, the other three to the subleading). Moreover, we should emphasize that we explicitly checked that the leading divergent term at $\lambda\to\infty$ (the first) is exact up to fourth order in the small-velocity expansion.

As briefly mentioned in footnote \ref{cutoffdefinition}, to enforce integration in the Coulomb region, we relate the cutoff $\lambda$ and $\r$ as $\lambda = \frac{\Lambda}{\r}$ with $\Lambda \gg 1$. The resulting metric 
is polyhomogeneous, of the form discussed in e.g. Ref.~\cite{Geiller:2024ryw}, with an overall power of $\frac{1}{\r^2}$ for the metric and $\frac{1}{\r^4}$ for the linearized Weyl tensor, with $\log \r$ enhancements.

To understand which terms may cancel between  Eqs.~\eqref{eq:metric_integral_subleading} and \eqref{eq:metric_integral_subleading} and their counterparts for $\kk>\lambda$, we explore how coefficients of the various tensor structures change in these equations under rescalings and shifts of the cutoff $\lambda$. 
To this end, we focus on the vector integral contributing to the metric, Eq.~\eqref{eq:metric_integral_subleading}. Using the relation explained in footnote \ref{cutoffdefinition} between $\lambda$ and $\r$, $\lambda = \frac{\Lambda}{\r}$ with $\Lambda \gg 1$ that enforces the restriction to the Coulomb region, it is easy to see that only terms proportional to $L^\mu$
can be changed by $\Lambda \rightarrow a\Lambda + b$ with fixed $a$ and $b$.
Thus, all terms proportional to $N^\mu$ are scheme-independent and should not be expected to be affected by radiation-region contributions that cancel the cutoff dependence. These are the terms contributing to $\Psi_0$ and $\Psi_1$ NP scalars.

A similar analysis of the two-tensor integral in Eq.~\eqref{eq:Weyl_integral_subleading} contributing to the Weyl tensor suggests a much stronger scheme dependence. However, since the linearized Weyl tensor is simply given by suitably antisymmetrized derivatives of the metric, its scheme dependence is substantially weaker than the tensor-integral analysis suggests. 
In particular, all terms that contain an $N^\mu$ coming from the metric are scheme-independent. The $N^\mu N^\nu$ structure has this property.

The exact expressions for the metric and NP scalars are rather unwieldy and not very illuminating. Their small velocity expansion, however, connects them to past results.
To this end, we parametrize the momenta of the incoming particles as 
\begin{equation}
    p_i^\mu = m_i u_i^\mu \equiv m_i (t^\mu + p_\infty v_i^\mu)\ ,
\end{equation}
and expand them in the limit $p_\infty \to 0$, corresponding to the nonrelativistic regime. In practice, this amounts to performing a post-Newtonian expansion. We find that the leading term in the expansion of the $\Psi_0$ NP scalar is 
\begin{equation}
    \Psi_0 = \frac{3 G^2 m_1 m_2}{2\r^4} \frac{(\varepsilon_+ \cdot v_1 - \varepsilon_+ \cdot v_2)^2}{\left[-(v_1-v_2)^2\right]^{3/2}} + \mathcal{O}(p_\infty^1)\ .
    \label{eq:finPsi0}
\end{equation}
This reproduces, up to an overall factor,
the result of Ref.~\cite{Christodoulou:2002}, demonstrating the breakdown of Sachs's peeling property in a two-body scattering process.
An analogous calculation, starting from the subleading soft theorem for an arbitrary number of incoming particles, will reproduce the generalization of the result to multi-particle scattering processes described in Ref.~\cite{Christodoulou:2002}.
Understanding the higher-order expansion of the 2-tensor integral in Eq.~\eqref{eq:metric_integral_subleading} and its rank-3 generalization will provide the generalization to arbitrary velocity of these peeling-violating contributions to the $\Psi_0$ NP scalar. 

We note that Eq.~\eqref{eq:finPsi0}, and the corresponding result of Ref.~\cite{Christodoulou:2002}, are different from the peeling violation predicted in Ref.~\cite{Damour:1985cm}. While both originate from the early-time ($t\to -\infty$) behavior of the quadrupole of the source,
\begin{equation}
    Q_{i j}(t) = A_{i j}\, t^2 - B_{i j}\, t \log(-t) + \dots ,
\end{equation}
the peeling violation of Ref.~\cite{Damour:1985cm} depends on the $A_{i j}$ and that of Ref.~\cite{Christodoulou:2002}  depends on $B_{i j}$.
We expect that the result of Ref.~\cite{Damour:1985cm} also originates 
from external gravitons in the Coulomb region. 
In particular, a dimensional analysis argument paralleling the one outlined at the beginning of this section suggests that a further source of peeling violation could be the subleading term in the soft-exchange region ($\sim \kk^0$), which we have not considered in our analysis. 
An alternative origin of the peeling violating terms of  Ref.~\cite{Damour:1985cm} could be terms in the Weyl tensor that are nonlinear in the metric fluctuations, in particular terms that depend on the linearized Schwarzschild metric of \emph{both} incoming particles.
We leave these for a future investigation.

The calculation leading to Eq.~\eqref{eq:finPsi0} provides the necessary ingredients for the evaluation of the $\Psi_1$ NP scalar. The result, to leading order in the PN expansion, is 
\begin{equation}
\label{eq:finPsi1}
\begin{split}
    \Psi_1 &= \frac{4 G^2 m_1 m_2}{\r^4} \frac{\left[\varepsilon_+\cdot(v_1 - v_2)\right] \left[(L+N)\cdot (v_1-v_2)\right]}{\left[-(v_1-v_2)^2\right]^{3/2}} \log(e^{-\gamma_E} \lambda)\\
    &+\frac{G^2 m_1 m_2}{4\r^4} \frac{\left[\varepsilon_+\cdot(v_1 - v_2)\right] \left[(55\, N - 81\, L)\cdot (v_1-v_2)\right]}{\left[-(v_1-v_2)^2\right]^{3/2}} \\
    &+ \frac{3 G^2 m_1 m_2}{\r^4} \frac{\left[\varepsilon_+\cdot(v_1 - v_2)\right] \left[(N- 3L)\cdot (v_1-v_2)\right]}{\left[-(v_1-v_2)^2\right]^{3/2}} \log(2 e^{+\gamma_E} \r) + \mathcal{O}(p_\infty^1)\ .
\end{split}
\end{equation}
Unlike $\Psi_0$, for which the breakdown of the peeling property involved a power-like enhancement, $\Psi_0\sim 1/\r^4$, over the expected behavior in Eq.~\eqref{eq:sachssmooth}, the $\Psi_1$ NP scalar is only logarithmically-enhanced. Such an enhancement was discussed for generic polyhomogeneous solutions of Einstein's equations in Ref.~\cite{Geiller:2024ryw}.

Before proceeding to the next section, let us briefly return to the scheme dependence of the contributions to $\Psi_0$ and $\Psi_1$ computed above.  
As we already explained, all contributions to the NP scalars coming from the terms proportional to the $N^\mu$ in the metric cannot be changed by adjusting the cutoff separating the Coulomb and radiation regions. The disappearance of the cutoff in Eq.~\eqref{eq:finPsi0} reinforces this conclusion.
A further indication that $\Psi_0$, as computed above, is scheme-independent will be given in the following section, where we will show that the radiation region cannot yield $\Psi_0 \sim 1/\r^4$ at the same order in $G$ as the Coulomb region.
$\Psi_1$, on the other hand, exhibits explicit cutoff dependence of the schematic form $\Psi_1 \sim G^2/\r^4\ln\lambda$. We expect that this dependence will cancel against divergences in the soft integration of the radiation region arising from logarithmic dependence on the retarded time.
Moreover, while we kept non-logarithmic terms in Eq.~\eqref{eq:finPsi1}, they should not be taken too seriously because they can be changed by altering the choice of cutoff in footnote~\ref{cutoffdefinition}.

\subsection{Radiation regions: the analytic contributions}
\label{sec:leadingwaveform}

Having analyzed the contribution of the Coulomb region to the NP scalars and recovered the departure from Sachs's peeling property found from different considerations in Refs.~\cite{Damour:1985cm,Christodoulou:2002},
we now proceed to discuss the contribution of the radiation region.
We will show that, in this region, the leading $1/\r$ behavior of the Newman--Penrose scalars is determined entirely by the on-shell metric. Moreover, the components of the momentum-space metric that are analytic for vanishing graviton virtuality, $k^{2}=0$, yield position-space contributions that satisfy Sachs's peeling property. The non-analytic terms will be examined in the next section.

As we discussed, the metric in Fourier space is given by the graviton form factor, with the subtraction of certain discontinuities. The external matter lines are on-shell and amputated through LSZ reduction, while the graviton line is off-shell and not amputated. 
Since we assume the current 
$J^{\mu\nu}$ is analytic at $k^2=0$ for the purpose of this section, the integral over $\omega$ 
can be evaluated using residue theorem. There are contributions
from the pole of the external graviton propagator, as well as from possible singularities in $\omega$ within the form factor. 
The LSZ reduction guarantees that such singularities are at $k^2\ne 0$. We will consider them separately and find that the pole of the external propagator contributes to the leading $1/\r$ falloff, while all the others lead to a faster (exponential) decay.

\subsubsection{The Fourier transform: retarded graviton poles}
\label{sec:leadingfourier}

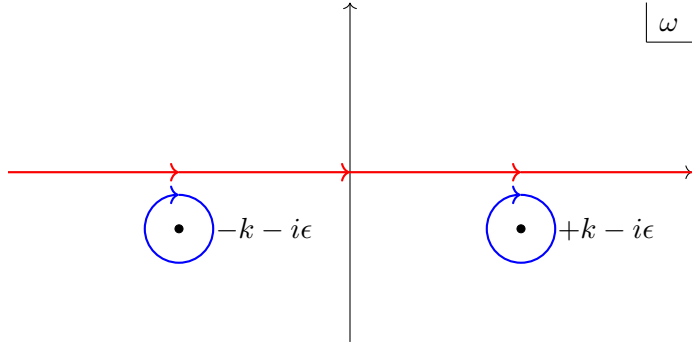
\begin{figure}[t]
    \centering
    \begin{tikzpicture}[scale=1.5]

        \draw[->] (-3,0) -- (3,0);
        \draw[->] (0,-1.5) -- (0,1.5);

        \node at (2.8,1.3) {$\omega$};
        \draw (2.6,1.5) -- (2.6,1.15); 
        \draw (2.6,1.15) -- (3,1.15); 

        \draw[thick, red, ->] (-3,0) -- (-1.5,0);
        \draw[thick, red, ->] (-1.5,0) -- (0,0);
        \draw[thick, red, ->] (0,0) -- (1.5,0);
        \draw[thick, red] (1.5,0) -- (3,0);

        \filldraw[black] (-1.5,-0.5) circle (1pt) node[right=10pt] {$-k - i\epsilon$};
        \draw[->, blue, thick] (-1.5,-0.2) arc[start angle=90, end angle=-270, radius=0.3];

        \filldraw[black] (1.5,-0.5) circle (1pt) node[right=10pt] {$+k - i\epsilon$};
        \draw[->, blue, thick] (1.5,-0.2) arc[start angle=90, end angle=-270, radius=0.3];

    \end{tikzpicture}
    \caption{At leading order, the current $J^{\mu\nu}(k)$ can be regarded to as an analytic function of $\omega$ and the integral over $\omega$ can be evaluated via residue theorem, on the poles introduced by the external (retarded) propagator. For $t  >0$, we can smoothly deform the original contour (in red) into the two contours encircling such poles (in blue).}
    \label{fig:treelevel}
\end{figure}

We begin by computing the contribution from the pole of the external propagator and we use the analyticity of the source to compute the $\omega$ integral; this effectively sets the source on-shell. Given Eq.~\eqref{eq:deformed_angular_contour_sec_3}, the result is 
\begin{equation}
    \label{eq:on_shell_waveform}
    \begin{split}
        \color{black}{h^{\mu\nu}(x) =} & \color{black}{+ \frac{1}{4\pi} \int_{0}^{\infty}\! \frac{d\kk}{2\pi} \, e^{+i \kk u} \kk^{d-3}\! \int\! d\hat{n}_{\perp} \left [ \int_{0}^{\infty}\! dy e^{-y \kk \r} (+2i y + y^2)^{\frac{d-4}{2}} \left. \htilde^{\mu\nu}(k) \right|_{\omega=-\kk} \right ]}\\
        & \color{blue}{- \frac{1}{4\pi} \int_{0}^{\infty}\! \frac{d\kk}{2\pi} \, e^{-i \kk v} \kk^{d-3}\! \int\! d\hat{n}_{\perp} \left [ \int_{0}^{\infty}\! dy e^{-y \kk \r} (+2i y + y^2)^{\frac{d-4}{2}} \left. \htilde^{\mu\nu}(k) \right|_{\omega=+\kk} \right ]}\\
        & \color{blue}{- \frac{1}{4\pi} \int_{0}^{\infty}\! \frac{d\kk}{2\pi} \, e^{+i \kk v} \kk^{d-3}\! \int\! d\hat{n}_{\perp} \left [ \int_{0}^{\infty}\! dy e^{-y \kk \r} (-2i y + y^2)^{\frac{d-4}{2}} \left. \htilde^{\mu\nu}(k) \right|_{\omega=-\kk} \right ]}     \\
        & \color{black}{+ \frac{1}{4\pi} \int_{0}^{\infty}\! \frac{d\kk}{2\pi} \, e^{-i \kk u} \kk^{d-3}\! \int\! d\hat{n}_{\perp} \left [ \int_{0}^{\infty}\! dy e^{-y \kk \r} (-2i y + y^2)^{\frac{d-4}{2}} \left. \htilde^{\mu\nu}(k) \right|_{\omega=+\kk} \right ]}\ ,\\
        =&\ \htilde_\text{ret}^{\mu\nu}(x) + \color{blue}{\htilde_\text{adv}^{\mu\nu}(x)} \ ,
    \end{split}
\end{equation}
where 
$u=t-\r$ and $v=t+\r$. The result is real because 
$ \htilde^{\mu\nu}(k)=\left(\htilde^{\mu\nu}(-k) \right)^*$. The first two terms originate from the two poles from the first term in Eq.~\eqref{eq:deformed_angular_contour_sec_3}, while the last two correspond to the second term in that equation. 
We discuss pairwise the four terms,
beginning with the first and the last term, which are complex conjugates of each other. 
In these terms, the Fourier conjugate of $\kk$ is the \textit{retarded time} $u$. Since we assume that $u$ is finite, the integral cannot be simplified without using properties of $\htilde^{\mu\nu}(k)$.
We can, however, understand their leading falloff with $\r$, which is determined by the angular integral over $y$.

Since we are in the radiation region ($k^\mu \sim \r^0$), we expand $\htilde^{\mu\nu}(k)$, or equivalently, the source $J^{\mu\nu}(k)$, in the angular variable $y$, around $y=0$.\footnote{The Watson's lemma guarantees that it is sufficient to consider the \textit{asymptotic} expansion of the source around such point.} A rescaling of $y$ turns this into a series expansion in $\r^{-1}$: each $y$ numerator factor pulls down a factor of $(\kk \r)^{-1}$.
The expansion in $y$ also generates a polynomial in $n_{\perp}$; the integral over these variables is carried out via identities like Eq.~\eqref{eq:perpintegraliden}.  
We are left with a non-trivial integral over $\kk$. 

The leading contribution corresponds to simply setting $y=0$ in Eq.~\eqref{eq:on_shell_waveform}:
\begin{eqnarray}
        h_\text{ret}^{\mu\nu}(x) & =& \frac{1}{4\pi} \int_{0}^{\infty}\! \frac{d\kk}{2\pi} \, e^{-i \kk u}\, \kk \!\int\! dn_{\perp} \left [ \int_{0}^{\infty}\! dy\, e^{-y \kk \r} \left. \htilde^{\mu\nu}(k) \right|_{\omega =\kk} \right ] + \text{c.c.} + \dots \ , \nonumber\\[5pt]
                      & \simeq& \frac{1}{4\pi \r} \int_{0}^{\infty}\! \frac{d\kk}{2\pi} e^{-i \kk u} \left. \htilde^{\mu\nu}(k) \right|_{\omega = \kk,y=0} + \text{c.c.} + \dots \ ,
    \label{eq:on_shell_waveform_4d}
\end{eqnarray}
where we also used the fact that, at $y=0$, the angular integration over $n_\perp$ is trivially 1, and the ellipsis stands for higher orders in $\r^{-1}$
We have thus recovered the KMOC result \cite{Cristofoli:2021vyo} for the leading $\r^{-1}$-contribution to the metric. 

We can also compute subleading contributions to the metric in the $\r^{-1}$ expansion by keeping higher-order terms in the expansion in $y$.
In the strict on-shell limit enforced by localizing onto the external propagator pole, the momentum-space metric $\tilde{h}^{\mu\nu}(k)$ becomes independent of $k^{2}$.
This suggests that we should not expect any departure from the leading-order behavior described above, at least if the initial state contains only two matter particles and no gravitational radiation.

\subsubsection{The Fourier transform: contribution of other poles}

Having understood the first and last terms in Eq.~\eqref{eq:on_shell_waveform} and their relation to the KMOC representation of the waveform, let us proceed to the
second and third terms in that equation.
They depend on the advanced time, which is parametrically large in the limit we are interested in, $v = t \,  +\r  \sim 2 \r \to \infty$. Thus the $\kk$ integral is dominated by $\kk \sim \r^{-1}$
which, together with the on-shell condition imposed by the external graviton pole, implies that
\begin{equation}
k^\mu \sim \r^{-1} \ .
\end{equation}
Thus, these integrals are dominated by graviton momenta in the Coulomb region, which we already discussed in Secs.~\ref{sec:leading_Coulomb} and \ref{Coulomb_ultra_soft}. In the radiation region, where $k^\mu \sim \r^{0}$, their contribution is exponentially suppressed by the Fourier phase factor.

We now discuss the contribution of other possible poles in $\omega$ that originate from the momentum-space metric $\htilde^{\mu\nu}(k)$. All such poles will set $\kk = a \omega$ where $a$ is some function of scalar products. This coefficient cannot be unity because 
because such a value also sets the external graviton on-shell and such a higher-order pole is in disagreement with known properties of 5-point S-matrix elements.
The residue of the Fourier-transform integrand at this pole is proportional to $\exp[\pm i \omega (t \pm a \r)]$. 
The exponent is parametrically large in all terms. Thus, as for the advanced poles, the contribution of these poles to the Fourier transform is from the Coulomb region, $k^\mu\sim \r^{-1}$. It is exponentially suppressed in the radiation region. 

While we presented the analysis of these poles' contributions 
to the metric, the same conclusion holds also for the Weyl tensor; the exponential suppression guarantees that such poles, even if present, do not spoil Sachs's peeling property.

\subsubsection{The NP scalars: analytic contributions}
\label{analytic}

Having understood that only the retarded poles of the external graviton propagator contribute in the radiation region, we now examine the asymptotic expansion for the  NP scalars. The gauge-invariant decomposition of the waveform Eq.~\eqref{eq:gauge_invariant_waveform} and the property of the coefficient functions 
$\alpha_i$ are regular at $z=1$ (or $y=0$) are essential in this analysis. In particular, this regularity property implies that these functions can be ignored in the $y$ integral as their non-constant parts contribute starting at subleading order in the $\r^{-1}$ expansion.

The tensor structure instead needs to be carefully analyzed. Indeed, as we are restricting ourselves to leading order in the $\frac{1}{\r}$ expansion of each NP scalar and considering the tetrad in the flat space limit, we can freely carry out the contraction inside the Fourier transform. Contractions with different combinations of the null tetrad with the tensor structure give different leading terms in the expansion around $z=1$. 

It is not difficult to see from Eq.~\eqref{eq:gauge_invariant_waveform} that for (the Fourier-transform of) $\Psi_{4}$, the expansion starts at order $(1-z)^0$:
\begin{equation}
    \Psitilde_{4}(k) = - (k\cdot N)^2\, \varepsilon_{-}^{\mu \nu} \htilde_{\mu \nu}(k) + \mathcal{O}((1-z)^1)\ .
\end{equation}
Moreover, we notice the following spinor version of the on-shell orthogonality condition of the Weyl tensor:
\begin{equation}
    \label{eq:Weyl_orthogonality_spinors}
    k^\mu \Ctilde_{\mu\nu\rho\sigma}(k) = 0 \quad \Leftrightarrow \quad \xi^{a} \Psitilde_{a b c d} = 0\ .
\end{equation}
This implies that, while the spinors $\lambda$ and $\rho$ are generally independent, they become linearly dependent when they are contracted with the self-dual part of the Weyl tensor. Indeed, using the definition of the spinor $\xi$ in terms of the basis spinors, Eq.~\eqref{eq:4D_k_on-shell}, leads to
\begin{equation}
    \rho^a \Psitilde_{a b c d} = - e^{+ i \phi} \sqrt{\frac{1-z}{1+z}} \lambda^a \Psitilde_{a b c d}\ .
\end{equation}
We can then use this relation to express all the momentum-space NP scalars in terms of $\Psitilde_4$, 
\begin{equation}
    \Psitilde_{j}(k) = e^{i (4-j) \phi - i \pi j} \left(\frac{1-z}{1+z}\right)^{\frac{4-j}{2}} \Psitilde_4 (k) \ ,
    \label{eq:psijitopsi4}
\end{equation}
establishing $\Psi_4 (k) $ as the generating function of all NP scalars at leading order in the $\r^{-1}$ expansion and for the contributions of the part of the metric that is analytic in the graviton frequency $\omega$.

To construct the position-space NP scalars we need to evaluate
\begin{equation}
    \Psi_{j}(x) = -\frac{1}{4\pi} \int_{0}^{\infty}\! \frac{d\omega}{2\pi} \, e^{-i \omega u}\, \omega \! \int_{0}^{2\pi}\! \frac{d\phi}{2\pi} \left [ \int_{0}^{\infty}\! dy\, e^{-y \omega \r} \Psitilde_{j}(k)  \right ] + \text{c.c.} + \dots \ .
    \label{eq:WeylTensor_position_space}
\end{equation}
In particular, the integral over the azimuthal angle vanishes unless the integrand is independent of $\phi$.
Since the angle $\phi$ enters\footnote{\label{foot:sh} A way to understand this dependence is from the structure of spherical harmonics.} only via the complex combination of coordinates $x\pm i y = e^{\pm i\phi}\sqrt{1-z^2}$ and, as discussed, the coefficient functions $\alpha_i$ are regular at $z=1$, the expansion of the NP scalar $\Psi_4$ around $z=1$ is
\begin{equation}
\label{eq:psi4rad}
    \Psitilde_4(k) = \sum_{m=-\infty}^{\infty} 
e^{i m \phi}\, (1-z^2)^{\frac{\abs{m}}{2}} \Psitilde^{(m)}_4(z)
\ ,
\end{equation}
where $\Psi^{(m)}_4(z)$ is nonvanishing and finite at $z=\pm1$. Thus, 
\begin{align}
\label{eq:expected_falloff}
    \int_0^{2\pi}\! \frac{d\phi}{2\pi}\, \Psitilde_j(k) \sim (1-z)^{4-j}
    ~~\Longrightarrow~~
    \Psi_j(x) \sim \int_0^\infty dy e^{-y \kk \r} y^{4-j}  \sim \frac{1}{\r^{5-j}}
    \ .
\end{align}

This concludes the main part of our proof that the part of the momentum-space metric that is analytic at $k^2=0$ yields a space-time metric that obeys Sachs's peeling property.

\subsubsection{Peeling violation vs. corrections to the tetrad} 
\label{sec:nonlinearitiesetc}

In our construction of the NP scalars, we used the tetrad in Eq.~\eqref{eq:metric_tetrad}. 
However, since the metric receives ${\cal O}(G)$ corrections, the tetrad may require corrections to satisfy its defining properties \eqref{eq:tetrad_props}.
Because of this, our analysis holds verbatim only at leading and next-to-leading orders in Newton's constant, and further analysis is necessary beyond these orders.
We argue below that such redefinitions of the tetrad cannot introduce any breakdown of peeling properties. 
In particular, the analytic part of the final-state graviton one-point function
discussed above will continue to exhibit Sachs's peeling properties. 

Parametrizing the tetrad in terms of two-component spinors as in Sec.~\ref{metricansatz}, all possible redefinitions are captured by
\begin{equation}
\rho\longrightarrow A(G) \rho + B(G) \lambda
\qquad
\lambda\longrightarrow C(G) \rho + D(G) \lambda \ ,
\label{spinorredef}
\end{equation}
for some coefficients $A, B, C, D$ so that $A(0) = 1 = D(0)$ and 
$B(0) = C(0) = 0$.  This redefinition maps the NP scalars in Eq.~\eqref{eq:NP_scalars} into linear combinations with coefficients that are quartic monomials in $A, B, C, D$. To study the effects of such transformations, it is sufficient to focus on the contributions to $\Psi_i$ that are proportinal to $\Psi_{j>i}$ because the latter have a slower fall-off than the former. 
For example,  $\Psi_0$ and $\Psi_1$ become
\begin{align}
&\!\!\!\!\!
\Psi_0 \rightarrow A(G)^4 \Psi_0 + 4A(G)^3 B(G) \Psi_1 + 6A(G)^2 B(G)^2 \Psi_2 + 4A(G) B(G)^3 \Psi_3 + B(G)^4 \Psi_4 
\\
&\!\!\!\!\!
\Psi_1 \rightarrow D(G) A(G)^3 \Psi_1 + 3D(G) A(G)^2 B(G) \Psi_2 + 3D(G) A(G) B(G)^2 \Psi_3 + D(G) B(G)^3 \Psi_4 \cr
&\!\!\!\!\!
~~~~~ + C(G) A(G)^3 \Psi_0 + 3C(G) A(G)^2 B(G) \Psi_1 + 3C(G) A(G) B(G)^2 \Psi_2 + C(G) B(G)^3 \Psi_3 \ .
\nonumber
\end{align}
 Each of the coefficients $B$ and $C$ is at least $\sim G/\r$ because 
 the first correction to the metric, corresponding to the direct sum of two noninteracting Schwarzschild black holes, is of this order. Moreover, we can use the leading values of the other two coefficients, $A=D = 1$. 
Thus, while the falloff of $\Psi_{i\ge 1}$ is slower than the expected falloff of $\Psi_0$, the fall-off of the coefficients $B$ and $C$ makes up for the difference. Consequently, the conclusion of Sec.~\ref{analytic} that the analytic contributions to the metric exhibit the peeling property are unaffected by redefinitions of the tetrad.
We expect that this holds beyond the leading order in the deformation, so that such mixing of NP scalars due to the nontriviallity of the nullness and orthogonality constraints on the tetrad does not, on its own, lead to departures from the peeling property of the Weyl tensor.

In this section, we focused on the part of the momentum-space metric that is analytic in $k^2$. Thus, the only singularity in this part of the metric originates in the outgoing-graviton propagator. Together with the second essential ingredient, the $z\rightarrow 1$ behavior of the coefficient functions $\alpha_{i}$ in Eq.~(\ref{eq:gauge_invariant_waveform}), it led to the position-space metric obeying Sachs's peeling property.
In the next section, we will discuss nonanalytic features of the final-state graviton one-point function and  their consequences on the large-distance fall-off of the NP scalars.


\subsection{Radiation region: non-analytic contributions and new peeling violation}
\label{sec:sublead}

The momentum-space metric exhibits non-analytic terms starting at ${\cal O}(G^3)$. In this section, we discuss the contributions to the NP scalars of the terms that, in the on-shell limit, yield the IR divergences of S-matrix elements~\cite{Weinberg:1965nx}.\footnote{\label{footnotenewdiver}Note that the waveform has another infrared divergence that is not regulated by taking the external graviton off-shell. This infrared divergence is related to how the time to closest approach of two separated bodies is logarithmically divergent as a function of initial separation in four dimensions \cite{Caron-Huot:2023vxl}. We regulate this other infrared divergence using dimensional regularization; it can also be regularized by taking the external matter particles off shell.}

We recall that the non-analytic terms can be constructed via the mapping identified in Sec.~\ref{sec:offshellgravitonregularization} at ${\cal O}(G^3)$, see Eq.~\eqref{eq:infrared_divergences}. In general, this mapping does not fully reconstruct the momentum-space metric. Indeed, the on-shell one-point function misses terms proportional to integer powers of the squared graviton momentum, $k^2$, and to $\varepsilon_{\mu\nu}(k) k^\nu$. The tensor factor in Eqs.~\eqref{eq:momentumspaceexpress}~and~\eqref{eq:projector} remedies the situation, producing a result proportional to $k^2$ when contracted with the graviton momentum. 
Moreover, any terms linear in $k^\mu$ that survive after this contraction drop out when evaluating the Weyl tensor and the NP scalars. Last, upon integration, these terms may contain non-analytic parts, but the $k^2$ factor brings an extra suppression in powers of $1/\r$ with respect to the non-analytic terms that arise from the on-shell mapping.

\subsubsection{New peeling violation from 
\texorpdfstring{${\cal O}(G^3)$}{O(G3)} contribution}
\label{sec:1looplogs}

\begin{figure}[t]
    \centering
    \begin{tikzpicture}[scale=1.5]

        \draw[->] (-3,0) -- (3,0);
        \draw[->] (0,-1.5) -- (0,1.5);

        \node at (2.8,1.3) {$\omega$};
        \draw (2.6,1.5) -- (2.6,1.15); 
        \draw (2.6,1.15) -- (3,1.15); 

        \draw[thick, red, ->] (-3,0) -- (-1.5,0);
        \draw[thick, red, ->] (-1.5,0) -- (0,0);
        \draw[thick, red, ->] (0,0) -- (1.5,0);
        \draw[thick, red] (1.5,0) -- (3,0);

        \filldraw[black] (1.2,-0.5) circle (1pt) node[right=10pt] {};
        \draw[->,decorate, decoration={snake},thick] (1.2,-0.5) -- (3,-0.5);
        \draw[blue, thick] (1.5,-0.65) arc[start angle=330, end angle=30, radius=0.3];
        \draw[-<, blue, thick] (1.5,-0.65)--(3,-0.65);
        \draw[->, blue, thick] (1.5,-0.35)--(3,-0.34);
        \filldraw[black] (1.2,-0.5) circle (1pt) node[below=11pt] {$\mathcal{C}_{1}$};

        \filldraw[black] (-1.2,-0.5) circle (1pt) node[right=10pt] {};
        \draw[->,decorate, decoration={snake},thick] (-1.2,-0.5) -- (-3,-0.5);
        \draw[blue, thick] (-1.5,-0.35) arc[start angle=150, end angle=-150, radius=0.3];
        \draw[->, blue, thick] (-1.5,-0.65)--(-3,-0.65);
        \draw[-<, blue, thick] (-1.5,-0.35)--(-3,-0.35);

        \filldraw[black] (-1.2,-0.5) circle (0pt) node[below=11pt] {$\mathcal{C}_{2}$};

    \end{tikzpicture}
    \caption{At ${\cal O}(G^3)$ the current $J^{\mu\nu}(k)$ has logarithmic branch-cuts as a function of $\omega$. We deform the initial integration contour along the real axis to the two contours ${\cal C}_{1,2}$ around the branch cuts.}
    \label{fig:looplevel}
\end{figure}
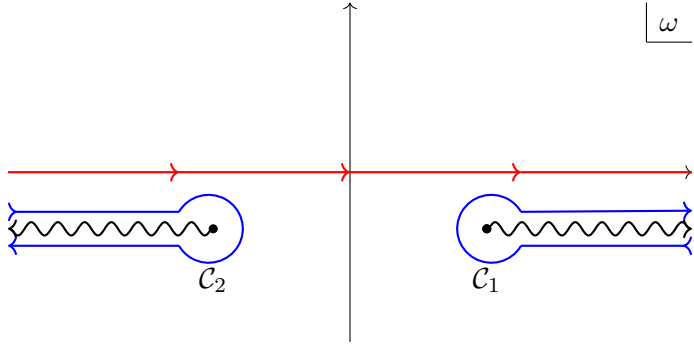

The starting point for the evaluation of the Fourier transform of the non-analytic part of the metric in Eq.~\eqref{eq:infrared_divergences} is Eq.~\eqref{eq:deformed_angular_contour_sec_3}. Due to the logarithmic branch cut,\footnote{We recall that, as discussed in Sec.~\ref{sec:offshellgravitonregularization} and in footnote~\ref{BMSfootnote}, logarithmic contributions to the final-state graviton one-point~function,~Eq.~\eqref{eq:h}, are both related to the tail effect and to the contribution of BMS modes at $\mathcal{I}^-$ (which in turn correspond to different configurations at finite time)}. the residues overlap with a branch point. Thus, instead of Fig.~\ref{fig:treelevel}, the contour for the $\omega$ integral is deformed as shown in Fig.~\ref{fig:looplevel}, leading to
\begin{align}
    \label{eq:deformed_angular_contour}
    &h^{\mu\nu}(x) \equiv h^{\mu\nu}(x)\big|_{{\cal C}_1} 
                         +h^{\mu\nu}(x)\big|_{{\cal C}_2}  \\
        & = \int_{\mathcal{C}_{1}+\mathcal{C}_{2}}\!\frac{d\omega}{2\pi i}\int_{0}^{\infty}\! \frac{d\kk}{2\pi}\frac{e^{-i\omega \, t-i\kk\r} \kk^{d-2}}{(\omega+i\epsilon)^{2}-\kk^{2}}\int\! dn_{\perp} \left [ \int_{0}^{\infty}\! \frac{dy}{2\pi} e^{-\kk y\r} (+2i y + y^2)^{\frac{d-4}{2}} \htilde^{\mu\nu}(k) \right ]   \nonumber \\
                        & -\int_{\mathcal{C}_{1}+\mathcal{C}_{2}}\!\frac{d\omega}{2\pi i}\int_{0}^{\infty}\! \frac{d\kk}{2\pi}\frac{e^{-i\omega \, t+i\kk\r} \kk^{d-2}}{(\omega+i\epsilon)^{2}-\kk^{2}}\int\! dn_{\perp} \left[ \int_{0}^{\infty}\! \frac{dy}{2\pi} e^{-\kk y\r} (-2i y + y^2)^{\frac{d-4}{2}} \htilde^{\mu\nu}(k) \right],
                        \nonumber
\end{align}
with ${\cal C}_{1,2}$ given in Fig.~\ref{fig:looplevel}. 

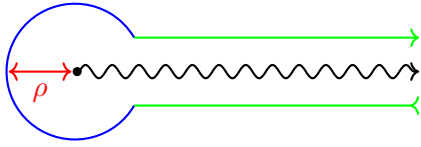
\begin{figure}[t]
    \centering
    \begin{tikzpicture}[scale=1.5]

        \draw[blue, thick] (1.5,-0.65) arc[start angle=330, end angle=30, radius=0.6];

        \filldraw[black] (1,-0.35) circle (1pt) node[right=10pt] {};
        \draw[->,decorate, decoration={snake},thick] (1,-0.35) -- (4,-0.35);

        \draw[-<, green, thick] (1.5,-0.65)--(4,-0.65);
        \draw[->, green, thick] (1.5,-0.05)--(4,-0.05);

        \draw[<->, red, thick] (0.95,-0.35)--(0.4,-0.35);
         \filldraw[black] (0.675,-0.35) circle (0pt) node[below=0.5pt] {\textcolor{red}{$\rho$}};

    \end{tikzpicture}
    \caption{We break up the integral over $\mathcal{C}_{1}$ into two parts, denoted in blue and green respectively. Both contours, integrated separately, are logarithmically divergent in $\rho$.}
    \label{fig:deform}
\end{figure}

We further break up the integral over each of the two contours into the contribution from the pole, given by the integral on a contour wrapping singularity, approaching the branch cut from both sides, with radius $\rho$ (eventually to be set to zero), and the contribution of the difference of the integrals above and below the cut, given by the discontinuity of the function. This decomposition is shown in Fig.~\ref{fig:deform} for ${\cal C}_1$.

To illustrate the evaluation, we discuss in some detail the integral on the third line of Eq.~\eqref{eq:deformed_angular_contour} on ${\cal C}_1$; we denote this contribution by $h^{\mu\nu}(x)\big|_{{\cal C}_1}^{(2)}$. The evaluation of the other integrals follows the same steps.
We have
\begin{align}
    \label{eq:deformed_w_contour2}
    &h^{\mu\nu}(x)\big|_{{\cal C}_1}^{(2)} = \\
        & + \int_{\textcolor{blue}{\rotatebox{270}{$\circlearrowright$}_\rho}}\!\frac{dw}{2\pi i}\int_{0}^{\infty}\! \frac{d\kk}{2\pi}\frac{e^{-i(w+\kk) \, t+i\kk\r} \kk^{d-2}}{(w+\kk+i\epsilon)^{2}-\kk^{2}}\int\! dn_{\perp} \left [ \int_{0}^{\infty}\! \frac{dy}{2\pi} e^{-\kk y\r} (-2i y + y^2)^{\frac{d-4}{2}} \htilde^{\mu\nu}(k) \right ] \notag \\ 
        & + \int_{\rho}^{\infty}\!\frac{dw}{2\pi i}\int_{0}^{\infty}\! \frac{d\kk}{2\pi}\frac{e^{-i(w+\kk) \, t+i\kk\r} \kk^{d-2}}{(w+\kk+i\epsilon)^{2}-\kk^{2}}\int\! dn_{\perp} \left [ \int_{0}^{\infty}\! \frac{dy}{2\pi} e^{-\kk y\r} (-2i y + y^2)^{\frac{d-4}{2}} \,\underset{w>0}{\textrm{Disc}}\, \htilde^{\mu\nu}(k) \right ] \notag  
    \end{align}
where we shifted the integration variable $\omega=w+\kk$. Both contributions in Eq.~\eqref{eq:deformed_w_contour2} are logarithmically divergent as $\rho\rightarrow 0$ because of the logarithmic terms in Eq.~\eqref{eq:infrared_divergences}; however, the dependence on $\rho$ cancels in the sum, leading to a divergence-free answer.
We are free to choose $\rho$ to be the smallest scale in the problem and perform the $w$ integration in a (asymptotic) small-$\rho$ expansion (with $\rho \ll {\r^{-1}}$).\footnote{The assumption that $\rho \ll \r^{-1}$ is consistent with the goal to evaluate the metric and NP scalars at higher orders in $\r^{-1}$. Also note that, even though $w\sim\rho \ll {\r^{-1}}$, $\omega\sim \kk\sim {\cal O}(\r^0)$, so we are still in the radiation region.}

Using the first term in Eq.~\eqref{eq:infrared_divergences} for 
$\htilde^{\mu\nu}(k)$, the first contribution is straightforward to compute because $\htilde_{\textrm{tree}}^{\mu}(k)$ is analytic as a function of $k^2$:
\begin{equation}
\begin{split}
\label{eq:IR_residue}
    &\int_{\textcolor{blue}{\rotatebox{270}{$\circlearrowright$}_\rho}}\!\frac{dw}{2\pi i} \, \frac{e^{-i w \, t}}{w (w + 2\kk)} \htilde^{\mu\nu}(k) \\
    &= i\frac{G}{\kk} (p_1+p_2)\cdot k\, \ln \left(\frac{2 \rho \kk}{\mu^2}\right) \left. \htilde_{\rm tree}^{\mu\nu}(k)\right|_{\omega=\kk} + \left.  \htilde_{\rm 1-loop}^{\mu\nu}(k)\right|_{\omega=\kk} +\mathcal{O}(\rho^1)\ .
    \end{split}
\end{equation}
The second requires some care in handling the rapidly oscillating exponential as $t  $ is taken large; the result is
\begin{align}
\label{eq:IR_disc}
        \int_{\rho}^{\infty}\!\frac{dw}{2\pi i} \frac{e^{-i w \, t}}{w (w + 2\kk)} \,\underset{w>0}{\textrm{Disc}}\, \htilde^{\mu\nu}(k) 
        & \simeq \left.  -i \frac{G}{\kk} \left[ \left(\gamma + \frac{i \pi}{2} + \log (\rho t \, ) \right) - \frac{i}{2\kk t \, }\right] (p_1+p_2)\cdot k \htilde_{\rm tree}^{\mu\nu}(k)\right|_{\omega=\kk} \notag \\
        & + \frac{G}{\kk t \, } \left[ \frac{\partial}{\partial \omega} (p_1+p_2)
        \cdot k \htilde_{\rm tree}^{\mu\nu}(k)\right]_{\omega=\kk}  + \dots \ ,
\end{align}
where the ellipsis stands for terms of higher orders in $\rho \ll {\r^{-1}} \ll \text{(any other scale)}$. We note that  $\log\rho$ cancels out between the pole and the discontinuity, Eqs.~\eqref{eq:IR_residue} and~\eqref{eq:IR_disc}, respectively. Moreover, the explicit $t  $ dependence can be recast 
as a dependence on $\r$ by writing 
\begin{align}
t \,  = \r+ u
\end{align}
and then expanded at large $\r$. Thus, the leading $r$ dependence is obtained simply by replacing $t  \rightarrow \r$.

Adding up the two contours in Fig.~\ref{fig:looplevel} and Eq.~\eqref{eq:deformed_angular_contour}, it follows that, at ${\cal O}(G^3)$ order, the terms proportional to 
$\ln(\r) \, \htilde^{\mu\nu}_{\textrm{tree}}(k)$ can be absorbed into a redefinition of the retarded time coordinate,  
\begin{equation}
T_r= t-\r-2G(m_{1}u_{1}\cdot n +m_{2}u_{2}\cdot n)\log(\mu \r) \ .
\end{equation}
In the next section, we will discuss the contribution of the all-order resummation of the $\log k^2$ dependence  to the graviton one-point function, and will note the emergence of the tortoise-like coordinate combination
\begin{align}
r_* = \r+2G(m_{1}u_{1}\cdot n +m_{2}u_{2}\cdot n)\log(\mu \r)
\end{align}
in the exponent of the one-dimensional Fourier transform from frequency to time domain. 

The second line of Eq.~(\ref{eq:IR_disc}) is governed by the leading off-shell properties of $\htilde_{\textrm{tree}}^{\mu\nu}(k)$, \textit{i.e.} the part of the graviton one-point function in which momenta are treated as off-shell while not being proportional to powers of $k^2$ or of the type $k^{(\mu} V^{\nu)}$ with some vector $V$. 
The latter is projected out when carrying out the analysis above directly for the Weyl tensor of for the NP scalars, while the former can contribute only subleading terms in the large-$\r$ expansion.

To compute the NP scalars we follow the same strategy as in previous sections, {\it i.e.} the metric tensor by the relevant contraction of the Weyl tensor and the null tetrad in the Fourier transforms. 
We have 
\begin{equation}
\label{eq:psi0defint}
\Psi_j=\Psi_j^{\perp}+\Psi_j^{\not\perp}
\end{equation}
where $\Psi_j^{\perp}$ denotes the contributions analogous to the first line of Eq.~\eqref{eq:IR_disc} (and originates from the part of the Weyl tensor that is transverse on $k^\mu$) and $\Psi_j^{\not\perp}$ denotes the analog of the second line of that equation.

We first consider $\Psi_0^{\perp}$. As we noted in the previous section, the transverse part of the Weyl tensor leads to the expected falloff, Eq.~\eqref{eq:expected_falloff}.
Evaluating the integral for the non-transverse part of the Weyl tensor, (the analog of the second line of \eqref{eq:IR_disc}) yields
\begin{align}
\label{eq:Psi0}
& \Psi_0^{ \not\perp}(x)  = \frac{G}{\r}\int \frac{d\kk}{2\pi} \kk^{d-3} e^{- i \kk u} \int d n_\perp
\int \frac{dy}{2\pi} e^{-\kk y \r}
\\
&\times \frac{\partial}{\partial\omega} \Big[(p_1+p_2)\cdot k \big(
(L\cdot k)^2 \varepsilon_+^\mu\varepsilon_+^\nu 
+
(\varepsilon_+\cdot k)^2 L^\mu L^\nu 
+
2(L\cdot k)(\varepsilon_+\cdot k) L^\mu \varepsilon_+^\nu 
\big) \; \htilde^\text{tree}_{\mu\nu}(k)\Big]\Big|_{\omega = \kk}
\nonumber \ .
\end{align} 
To evaluate the remaining integrals it is convenient to parametrize the graviton momentum in terms of the null tetrad: 
\begin{align}
    \label{eq:parameterizationk_4d}
        k^\mu & = \frac{\omega}{\sqrt{2}} \left[(1+z) L^\mu + (1-z) N^\mu + \sqrt{1-z^2} \cos\phi \,(\varepsilon_{+}^\mu + \varepsilon_{-}^\mu) + i \sqrt{1-z^2} \sin\phi \,(\varepsilon_{+}^\mu - \varepsilon_{-}^\mu) \right] \notag \\
              & = \frac{\omega}{\sqrt{2}} \left[(1+z) L^\mu + (1-z) N^\mu + \sqrt{1-z^2} \left(e^{i \phi} \varepsilon_{+}^\mu + e^{-i \phi} \varepsilon_{-}^\mu \right) \right]\ .
\end{align}
together with the properties of the null tetrad, Eq.~\eqref{eq:tetrad_props}, it implies that 
\begin{align}
\label{tetrad_products}
L\cdot k = \frac{\omega}{\sqrt{2}} (1-z)
\qquad 
\varepsilon_+\cdot k = -\frac{\omega}{\sqrt{2}}e^{-i\phi}\sqrt{1-z^2} \ .
\end{align}
The three terms in Eq.~\eqref{eq:Psi0} have different $\phi$ dependence, and the integral over $\phi$ vanishes unless each factor of $e^{i\phi}$ pulls its conjugate out of $\partial_\omega((p_1+p_2)\cdot k\, \htilde^{\mu\nu}_\text{tree}(k))$. As in the previous section, such term comes with at last one factor of $\sqrt{1-z^2}$. 
Overall, each of the three terms starts out as $(1-z)^2 \sim y^2$ in a small-$y$ expansion and thus yields a factor of $\r^{-3}$. Together with the overall factor in Eq.~\eqref{eq:Psi0} it follows that
\begin{equation}
\label{eq:leadingpeelingviolation}
     \Psi_0 (x) \sim \frac{1}{\r^4}\ .
\end{equation}
Thus, we find that the large-distance scaling of the radiation contribution to $\Psi_0$ at ${\cal O}(G^3)$ departs from Sachs's peeling property. This departure is, however, for a different reason than the Coulomb region.\footnote{We also recall that, as we discussed in Sec.~\ref{Coulomb_ultra_soft}, the Coulomb-region contributions are confined to the tree-level metric while the contribution found here is at a higher order.}
This departure is due to the non-analyticity of the spacetime metric from the long-range nature of gravitational interactions, which bring off-shell contributions of the graviton one-point function in the large-$\r$ expansion.\footnote{We note that, since it is entirely due to outgoing gravitons, the same phenomena should also occur in the bound case. See Ref. \cite{Ivanov:2025ozg} for a discussion of tails in the bound and unbound context. On the other hand, the same strategy cannot be straightforwardly applied to the bound case, as it is not possible to define an asymptotic bound state for macroscopic gravitational systems (strictly speaking, bound states in relativistic classical mechanics do not exist due to radiation emission).}

It is straightforward to follow the same strategy and evaluate $\Psi_{1}$. As for $\Psi_0$, the transverse part $\Psi_1^\perp$ exhibits a falloff  consistent with Sachs's peeling property, while the nontransverse part $\Psi_1^{\not\perp}$ departs from it. The relevant integrand is
\begin{align}
\label{eq:Psi1}
\Psi_1^{ \not\perp}(x)  &= \frac{G}{\r}\int \frac{d\kk}{2\pi} \kk^{d-3} e^{- i u \kk} \int d n_\perp
\int \frac{dy}{2\pi} e^{-\kk y \r}
\frac{\partial}{\partial\omega} \Big[(p_1+p_2)\cdot k\, \big(
(L\cdot k)^2 \varepsilon_+^{\mu} N^{\nu} 
\\
&\quad
+
\frac{1}{2}(\varepsilon_+\cdot k)(N\cdot k) L^\mu L^\nu 
+
2(L\cdot k)(\varepsilon_+\cdot k) L^\mu N^\nu +(\varepsilon_+\leftrightarrow N)
\big) \; \htilde^\text{tree}_{\mu\nu}(k)\Big]\Big|_{\omega = \kk}
\nonumber \ .
\end{align} 
Using Eq.~\eqref{tetrad_products}, $N\cdot k = \frac{\omega}{\sqrt{2}} (1+z)$ and the properties of spherical harmonics (see footnote~\ref{foot:sh} and the discussion around Eq.~\eqref{eq:psi4rad}), it is not difficult to see that the first term on the second line of Eq.~\eqref{eq:Psi1} and the image of the second term under the map $(\varepsilon_+\leftrightarrow N)$
imply that at ${\cal O}(G^3)$ scales as 
\begin{equation}
\label{eq:leadingpeelingviolationpsi1}
\Psi^{\not\perp}_1 (x) \sim \frac{1}{\r^3}\ .
\end{equation}
This constitutes a stronger violation of the peeling property than that found in the Coulomb region. 
For the same reason, the Coulomb-region contribution cannot cancel this.
Eq. (\ref{eq:leadingpeelingviolationpsi1}) is one of our main technical results.

The scaling of $\Psi_0$ and $\Psi_1$ at large $\r$ were given by simple dimensional arguments and the inspection of the angular dependence of the momentum-space Weyl tensor. These arguments are not sufficient to fix the scaling of $\Psi_2$, which is sensitive to terms $\propto k^2$ in the metric. An explicit computation of these contributions is needed to understand the contribution of the tails to the falloff of this NP scalar.

As this point, we should carefully provide an analysis similar to the one in Sec.~\ref{sec:leadingwaveform} regarding the effects of possible corrections to the tetrad. Since the corrections to the tetrad are expected to be ${\cal O}(G/\r)$ and ${\cal O}(G^2/\r)$ (which is the order of the correction to the metric itself), they will mix with the leading and next-to-leading terms in the NP scalars. Such terms preserve peeling. Thus, the argument presented in Sec.~\ref{sec:nonlinearitiesetc} applies identically here. We also analyze the possible contribution from the non-linearities of the Weyl tensor on the metric. Indeed, at quadratic order in $h_{\mu\nu} = g_{\mu\nu}-\eta_{\mu\nu}$ the Weyl tensor is given by 
\begin{equation}\label{eq:quadraticweyl}
\begin{split}
C{}_{\rho\sigma \mu \nu } & ={\cal O}(h_{\mu\nu})+ \eta^{\lambda\zeta}\Gamma_{\lambda \rho \mu }\Gamma_{\zeta\nu \sigma }-\eta^{\lambda\zeta}\Gamma_{\lambda \rho \nu}\Gamma_{\zeta \mu \sigma } + {\cal O}(h_{\mu\nu}^3)\ ,\\
&\textrm{where}\quad 
\Gamma_{\rho \nu \lambda }=\frac{1}{2}\left ( {\partial_{\lambda}}h_{\rho\nu}+{\partial_{\nu}}h_{\rho\lambda}-{\partial_{\rho}}h_{\nu\lambda}\right ) \ . \\
\end{split}
\end{equation}
The NP scalars pick up the corresponding projections of these terms on the null tetrad. One may wonder if such contributions can alter the conclusion that non-analytic contributions to the metric are a novel source of peeling violation at ${\cal O}(G^3)$. We will argue here that this cannot be the case.

Since we are estimating the bilinear contributions to the linearized Weyl tensor at a fixed order in $G$, the two metric fluctuations must be of lower order. At ${\cal O}(G^3)$ they can only be either the expanded Schwarzschild metric or the tree-level scattering metric. 
Each factor has an ${\cal O}(1/\r)$ falloff, see {\it e.g.} Sec.~\ref{sec:leadingfourier}.
Direct calculations (or dimensional analysis) indicate that each derivative of the (expanded) Schwarzschild metric is suppressed by ${\cal O}(1/\r)$ relative to the metric.  
Recalling the form of the tree-level scattering metric as a Fourier transform, it follows that its derivative (or the corresponding momentum) points along a null direction; at subleading orders there is a departure from the light-like direction at least of ${\cal O}(1/\r)$. 
Thus, the two-fluctuation part of the Weyl tensor falls off at least like ${\cal O}(1/\r^3)$ because in \eqref{eq:quadraticweyl} at least one derivative acts on a Schwarzschild metric and the other derivative points in a null direction up to ${\cal O}(1/\r)$ corrections. 
To complete the argument, we must inspect the effects of the projection of the Weyl tensor onto the null tetrad. 
The four free indices of the Weyl tensor in \eqref{eq:quadraticweyl} can be supplied either by the background Minkowski metric or, as dictated by soft-region power counting, by a pair of graviton momenta.
The null nature of the tetrad projects out the contributions bilinear in the Minkowski metric for both $\Psi_0$ and $\Psi_1$.
In the remaining terms, two tetrad vectors are contracted with 
derivatives on a metric fluctuation -- either of the expanded Schwarzschild 
metric or of the tree-level scattering metric. 
Since both of them are at an order in $G$ at which the peeling property is observed, we can infer that the scalar product of $L$ and $\varepsilon_{+}$ vectors and a derivative falls off at least as ${\cal O}(1/\r)$.\footnote{We already discussed their properties explicitly in Eq.~\eqref{tetrad_products} and below Eq.~\eqref{eq:Psi1}.}
Since for $\Psi_0$ both derivatives are contracted with either $L$ and $\varepsilon_{+}$, and for $\Psi_1$ at most one derivative is contracted with $N$, see Eqs.~\eqref{eq:NPscalars}, we conclude that the bilinear corrections to the NP scalars fall off as 
\begin{equation}
\Psi_0\Big|_{\text{bilinear}} \sim {\cal O}(
1/\r^5)
\qquad
\Psi_1\Big|_{\text{bilinear}} \sim {\cal O}(
1/\r^4) \ .
\end{equation}
That is, bilinear corrections do not change the conclusion reached in  Eqs.~\eqref{eq:leadingpeelingviolation} and \eqref{eq:leadingpeelingviolationpsi1} that $\Psi_0$ and $\Psi_1$ fall off as $\r^{-4}$ and $\r^{-3}$, respectively.
%

\subsubsection{Fourier transform with exponentiated logarithms}
\label{sec:exponentiated_FT}

A similar analysis can be carried out for the exponentiated IR divergences regularized by the off-shell graviton momentum:
\begin{equation}
\begin{split}
    \label{eq:exponentiated_log}
    \htilde^{\mu\nu}(k,\epsilon)     &= \left(\frac{-4\pi k^2}{\mu^2} \right)^{-2i G (\omega t + \kk n)\cdot (p_1+p_2)}\, \widetilde h_0^{\mu\nu}(k, \epsilon)\ ,
    \end{split}
\end{equation}
where $\widetilde h_0^{\mu\nu}(k,\epsilon)$ is finite at $k^2=0$, but it is not IR finite in general. Indeed, in the case of scattering, $\widetilde h_0^{\mu\nu}(k, \epsilon)$ has extra IR divergences associated to the logarithmic drift of the worldlines of the incoming particles 
at early times \cite{Caron-Huot:2023vxl}. 

Since Eq.~\eqref{eq:exponentiated_log} is non-analytic in $k^2=0$, the calculation follows the same steps as in Sec.~\ref{sec:1looplogs}.
It is not difficult to notice that the 
phase in Eq.~\eqref{eq:exponentiated_log} regulates the integration along the discontinuity of the $\omega$ integral in Eq.~\eqref{eq:deformed_w_contour2}. 
This implies that there is no need to separately consider the contours wrapping the residues and the branch cuts because the former can be discarded as we shrink the radius of the contour to zero. Thus, the only contribution to the Fourier transform comes from the branch cut: 
\begin{equation}
    \begin{split}
        D &= \int_{0}^{\infty}\!\frac{dw}{2\pi i}\, \frac{e^{-i w \, t}}{w (w+2\kk)} \underset{w>0}{\textrm{Disc}}\, \htilde^{\mu\nu}(k, \epsilon)\\
        &= \int_{0}^{\infty}\!\frac{dw}{2\pi}\, e^{- w \, t}\, \frac{\sin\left(2i \pi G {\cal E}^\prime \kk + 2 \pi G ({\cal E}^\prime - {\cal E}^{\prime \prime}) w \right)}{\pi \left[- w (w+2i \kk)\right]^{1+2i G {\cal E}^\prime\kk - 2G({\cal E}^\prime - {\cal E}^{\prime \prime}) w}} \left. \widetilde h_0^{\mu\nu}(k, \epsilon) \right|_{\omega=\kk-i w}\ ,
    \end{split}
\end{equation}
where we defined a shorthand notation:
\begin{equation}
    \begin{split}
        \mathcal{E}^\prime &= (t + n) \cdot (p_1 + p_2)\ ,\\
        \mathcal{E}^{\prime \prime} &= n \cdot (p_1 + p_2)\ .
    \end{split}
\end{equation}
Because of the exponential damping, the integral is dominanted by $w\sim t \, {}^{-1}$, so we can evaluate this integral as an asymptotic expansion at large $t$: 
\begin{align}
        D &\sim \left(-\frac{2 i \kk}{\mu^2\, t }\right)^{-2i G {\cal E}^\prime \kk} \frac{\sin\left(-2i \pi G {\cal E}^\prime\kk\right) }{\pi}\Gamma\left[-2i G {\cal E}^\prime\kk\right] \left\{- \frac{1}{2\kk} \left[1 + \frac{2 G {\cal E}^{\prime \prime}}{t} H(t,\kk) \right] {\left. \widetilde h_0^{\mu\nu}(k, \epsilon)\right|_{\omega=\kk}}\right. \nonumber \\
        & \left.+ \frac{2G {\cal E}^\prime\kk}{(2 \kk)^2 \, t } \left[ \left(2 H(t,\kk) - 2i G{\cal E}^\prime\kk-1\right) \widetilde h_0^{\mu\nu}(k, \epsilon)+ 2 \kk \frac{\partial}{\partial \omega} \widetilde h_0^{\mu\nu}(k, \epsilon)\right]_{\omega=\kk} + \dots \right\}\ ,
\end{align}
where
\begin{equation}
    \begin{split}
        H(t, \kk) &= 2 i G \mathcal{E} \kk \left[i \pi  \coth (2 \pi  G \mathcal{E}^\prime \kk)+\psi ^{(0)}(1-2 i G \mathcal{E}^\prime \kk)+\log \left(-\frac{2 i \kk}{\mu^2 t}\right)\right]\ ,
    \end{split}
\end{equation}
and we have not yet evaluated the angular integration over $n^\mu$.

This expression clearly shows how the exponentiation of the IR divergences implies the exponentiation of the $\ln \mu \r$ contributions (upon writing the time in terms of the retarded time and expanding in $u/\r$).\footnote{To write the entire prefactor as a exponential, one may use the identity:
\begin{equation}
    \frac{1}{\pi} \sin(\pi a) \Gamma[a] = e^{- \gamma_E a - \sum_{n=2}^\infty \frac{\zeta (n)}{n} a^n}\ ,
\end{equation}
where $\zeta(n)$ is the Riemann zeta function.
}
It also indicates that 
\begin{equation}
T_{r}= t - (\r+2G(m_{1}u_{1}\cdot n +m_{2}u_{2}\cdot n)\log(\mu \r)) \ .
\end{equation}
is the appropriate variable conjugate to the frequency $\kk$, 
in the Fourier transform to time domain.

The core of the analysis of the scaling of the NP scalars relies only on the properties of the tetrad {\it vis \`a vis} the graviton momentum and the properties of spherical harmonics. These are unaffected by the resummation of off-shell-regularized IR divergences, leading to the same conclusion as in Sec.~\ref{sec:1looplogs}, {\it i.e.} that the large-distance scaling of $\Psi_0$ and $\Psi_1$ departs from those impled by asymptotic simplicity.

\section{Conclusions}
\label{sec:conclusions}

In this paper, we studied the structure of spacetime near null infinity using methods from the amplitudes program and asymptotic expansions. We computed the leading and sub-leading contributions to the Penrose scalars in the PM expansion, reproducing violations of Sachs's peeling theorem previously predicted from different considerations in Refs.~\cite{Damour:1985cm,Christodoulou:2002,Winicour:1985}. 
In turn, this implies a departure from Penrose's proposal of asymptotic simplicity~\cite{Penrose:1965am}. 

In addition to recovering results of Refs.~\cite{Damour:1985cm,Christodoulou:2002,Winicour:1985}, we also
found a novel source of violation of the peeling property from nonlinear, long-range interactions between localized sources and the surrounding gravitational field. 
These lead to stronger departures from Penrose's asymptotic simplicity proposal than previously known. 
An important aspect of our work is identifying the contributing regions of the external-graviton momentum integral and they interact with the regions of the integration domain of the exchanged momentum.

We expect these results to be relevant for flat-space analyses that use the metric's asymptotic behavior. For example, the asymptotic structure of spacetime is essential for the presence of asymptotic symmetries.
BMS symmetries are known to persist under the milder violations of peeling~\cite{Geiller:2024ryw, Fuentealba:2024lll} that follow from the universal the soft theorems. Whether they also survive the more pronounced violations identified in this work, and consequently the identification of canonical notions of global mass and angular momentum~\cite{Flanagan:2015pxa}, is a question for separate study.\footnote{Aidan Herderschee thanks Gautam Satishchandran for discussion on this topic.} 

We contrast here two complementary observations. 
On the one hand, the transformation of the NP scalars under general coordinates transformations, large or small, suggests that their leading order in the $1/\r$ expansion cannot be changed by diffeomorphisms that preserves the asymptotic-Minkowski nature of the spacetime. In particular, the leading-order peeling violation cannot be removed by such transformations. 
On the other, a BMS transformation with parameter twice that of Veneziano and Vilkovisky~\cite{Veneziano:2022zwh} flips the sign of the disconnected-cut contribution to the waveform~\cite{Bini:2024rsy,Elkhidir:2024izo} and, in an off-shell regularization~\cite{Elkhidir:2024izo}, appears to completely cancel the tail-related logarithmic terms in Eq.~\eqref{eq:infrared_divergences_all}, $iG (p_1\cdot k \ln (u_1\cdot k)^2/k^2) + p_2\cdot k \ln (u_2\cdot k)^2/k^2))$.
In this BMS frame, which correspond to a very specific choice of boundary conditions at finite time and is distinct from both the canonical and the intrinsic ones, the novel peeling violation that we identified in~Sec~\ref{sec:sublead} appears to cancel out. 
%
From our analysis, it is unclear why such choice of boundary conditions are special and we leave further investigations for future work. In particular, it would be interesting to explore the precise map relating different BMS frames at $\mathcal{I}^-$ to boundary conditions at space-like hypersurfaces at $t\to - \infty$.
%

It would be fascinating if the subleading components of the metric in $\r^{-1}$ were experimentally measurable. Naively, these effects are enormously suppressed by the vast distance between astrophysical events and gravitational-wave observatories, making them largely formal considerations. Nevertheless, the computational techniques developed in this paper could prove useful in contexts where finite-distance effects become relevant, such as in comparisons with numerical relativity simulations~\cite{Mitman:2024uss} and at colliders~\cite{Collins:2019ozc}.

The framework developed in this paper can be used to compute the scattering metric throughout spacetime. This calculation amounts to evaluating exactly the Fourier transform over the external-graviton momentum of the final-state graviton one-point function.
Assuming that such an exact metric were available, it should be possible to direct test of asymptotic simplicity by computing the conformal structure and demonstrating that they are not regular at null infinity. 

\acknowledgments

We thank Martin Beneke, Donato Bini, Thibault Damour, Holmfridur Hannesdottir, Carlo Heissenberg, Johannes Henn, Enrico Herrmann, Gary Horowitz, Gregory Korchemsky, Yao Ma, Sebastian Mizera, Michael Ruf, Lorenzo Tancredi, Gautam Satishchandran, George Sterman, Robert Szafron, and C\'eline Zwikel for insightful discussions.  AH is grateful to the Simons Foundation as well as the Edward and Kiyomi Baird Founders' Circle Member Recognition for their support. SDA was supported by the European Research Council, under grant ERC-AdG-885414. R.R. is supported by the U.S.  Department of Energy (DOE) under award number~DE-SC00019066 and by a Senior Fellowship at the Institute for Theoretical Studies, ETH Z\"urich. F.T. is supported by the startup grant from the department of physics at Fudan University.
Part of the work reported in this paper was carried out while R.R. and F.T. participated in the ``What is Particle Theory?'' program at the Kavli Institute for Theoretical Physics (KITP) and supported in part by grant NSF PHY-2309135.

\bibliographystyle{JHEP}
\bibliography{Draft.bib}

\end{document}